\documentclass[11pt]{article}
\usepackage{amsmath,amsfonts,amssymb}
\usepackage{graphicx,color}
\usepackage{epsfig,epsf}
\usepackage[toc,page]{appendix}
\usepackage{bm}
\usepackage{dcolumn,subfig}
\usepackage{array}
\usepackage{cite}
\usepackage{color}

\usepackage{relsize}
\def\Babar{{\mbox{\slshape B\kern-0.1em{\smaller A}\kern-0.1em B\kern-0.1em{\smaller A\kern-0.2em R}}}}

\usepackage[bookmarks, breaklinks, colorlinks,urlcolor=black, citecolor=red, 
linkcolor=blue]{hyperref}

\usepackage{booktabs}
\usepackage{graphicx}
\graphicspath{ {images/} }
\usepackage{amssymb}

\usepackage{graphicx}
\usepackage{graphics}
\DeclareGraphicsExtensions{.eps}
\usepackage{float} 
\usepackage{mathrsfs}
\usepackage{amsmath}
\usepackage{slashed}
\usepackage{amsmath}
\usepackage{amsfonts}   
\usepackage{amssymb}
\usepackage{subfloat}

 \usepackage{color}

 \usepackage[normalem]{ulem}

 \definecolor{darkgreen}{cmyk}{1,0,1,0.2}
 
\def\com2#1{\textcolor{red}{\it{#1}}}

\def\be {\begin{equation}}
\def\ee {\end{equation}}
\def\bea {\begin{eqnarray}}
\def\eea {\end{eqnarray}}
\def\n {\nonumber}
\def\bra {\langle}
\def\ket {\rangle}

\def\mzt {\mathbb{Z}_2}

\begin{document}

\renewcommand*{\thefootnote}{\fnsymbol{footnote}}

\begin{center}
 {\Large\bf{A 96 GeV scalar tagged to dark matter models}}\\[5mm]
 {\bf Anirban Kundu}\footnote{anirban.kundu.cu@gmail.com},
{\bf Suvam Maharana}\footnote{msuvam221@gmail.com}, and 
{\bf Poulami Mondal}\footnote{poulami.mondal1994@gmail.com}\\[3mm]
Department of Physics, University of Calcutta, \\
92 Acharya Prafulla Chandra Road, Kolkata 700009, India
 \\ 
 \today
 \end{center}


\begin{abstract}

Recently, the CMS Collaboration observed the hint of a resonance decaying to two photons at about
96 GeV with a local significance of $2.8\sigma$. While it is too early to say whether this will stand the test of
time, such a resonance can easily be accommodated in many extensions of the Standard Model (SM). 
The more challenging part is to tune such an extension so that the required number of diphoton events is 
reproduced. Assuming that the new resonance is a scalar, we propose that the signal may come either from 
an ultraviolet complete model with vectorial quarks, or a model involving gluon-scalar and photon-scalar 
effective operators. We then incorporate this portal to several extensions of the SM that include one or more 
cold dark matter candidates, and try to investigate how the existence of such a scalar resonance affects the 
parameter space of such models. As expected, we find that with such a scalar, the parameter space 
gets more constrained and hence, more tractable. We 
show how significant constraints can be placed on the parameter space, not only from direct dark matter 
searches or LHC data but also from theoretical considerations like scattering unitarity or stability of the potential,
and discuss some novel features of the allowed parameter space.

\end{abstract}



\setcounter{footnote}{0}
\renewcommand*{\thefootnote}{\arabic{footnote}}

\tableofcontents

\section{Introduction}

The existence of cold dark matter (CDM) is perhaps the biggest motivation to search for physics beyond the 
Standard Model (SM). The SM does not contain any suitable CDM candidate, something that is massive,
and singlet under $U(1)_{\rm em}$ and $SU(3)_{\rm c}$. Any extension of the SM with a $\mzt$ symmetry
for which all the SM fields are $\mzt$-even can potentially come to the rescue, with the CDM candidate being
the lightest $\mzt$-odd particle. The lightest neutralino of the R-parity conserving supersymmetric models, 
and the lightest Kaluza-Klein particle of the Universal Extra Dimension models, fall in this category. 
Gauge singlet objects like massive right-handed neutrinos, or additional scalars, can also fit the bill. 
In our subsequent discussion, we will assume a thermalised non-baryonic CDM with the density given 
by \cite{pdg2018}
\be
\Omega_{\rm CDM} h^2 = 0.1186 \pm 0.0020\,,
\label{eq:omegadm}
\ee
where $h = H_0/100$ is the reduced Hubble constant\footnote{Not to be confused with the 125 GeV scalar resonance, 
which we will also denote by $h$.}. The existing literature is abound with the possible 
models of CDM, their detection strategies, and other phenomenological aspects. Let us refer the 
reader, in general, to some review articles \cite{bertone}, in particular, to Ref.\ \cite{1903.03616}
for the analysis of several Higgs portal CDM models, and to Ref.\ \cite{1703.07364} for their collider signatures.

It is always an interesting exercise to link the CDM with some other beyond-SM signals or even motivations, like 
the generation of neutrino masses, or the flavour anomalies. In this paper, we will study some CDM models 
in conjunction with the recently observed hint of a (possibly) scalar resonance, 
close to $2.8\sigma$ or 98\% CL, at about 96 GeV from the CMS 
Collaboration \cite{1811.08459}, decaying to a diphoton final state, with $\sim 370\, (1070)$ events in the 8 (13)
TeV dataset.  Although there has been neither confirmation nor denial from the ATLAS Collaboration, this
hint has given rise to a lot of interesting theoretical speculations, and the physics implication of such a resonance 
in the context of beyond-SM scenarios, including its collider signatures, has been discussed in Refs.\ 
\cite{1612.08522,1812.00107,1812.05864,1905.03280,1906.03389}. The astrophysics implications of models with a light scalar decaying into two photons have been studied in Ref.\ \cite{1906.02175}. 

How does this new scalar, which we will call $\chi$, help in the analysis? 
While $\chi$---assuming that it exists---can never be a CDM candidate, it can very well be one component of 
a multiplet with the other component(s) constituting the dark matter, or it can be a mediator for the CDM-SM 
interaction. A fairly straightforward example, which we will
discuss in detail, is to add a complex singlet to the SM; one of its components is the DM, while the other leads to
$\chi$. The interesting part is that such a scenario reduces the number of free parameters in the model, 
and hence the allowed parameter space for the CDM becomes more tractable. This is, in a sense, the 
rationale of this paper. 

In this paper, we consider three models that accommodate both $\chi$ and a prospective 
CDM candidate. They are, respectively,
\begin{enumerate}
\item Complex Scalar (CS): This consists of the SM augmented by a complex scalar singlet $S$. One of the 
components of $S$ mixes with the 
SM doublet, and these two states appear as $\chi$ and $h$, the 125 GeV Higgs boson\footnote{The 
Higgs portal of real scalar singlet CDM is already ruled out, except either in a narrow resonance region
around $m_h/2$, or for large CDM mass.}. We further refer the reader to Refs.\ 
\cite{1906.02175,1202.1316,1309.6632,1812.05952}
for more discussion on the CDM component of a complex scalar augmented SM. 

\item Real Scalar with Fermion (RSF): Here one has a real gauge singlet scalar (which mixes with the SM 
doublet), and a vectorial gauge singlet fermion $\psi$ that plays the role of the CDM candidate.

\item Complex Scalar with Fermion (CSF): This is, in a sense, an amalgamation of the previous two models. 
We take a complex singlet scalar $S$ with the singlet fermion $\psi$, and consider the possibility where 
we have two CDM candidates, one scalar and the other fermionic. For such singlet fermionic CDM models,
we refer the reader to, {\em e.g.}, Refs.\ 
\cite{0708.3790,0803.2932,1112.1847,1305.3452,1407.1859,1806.08080}.
The two-component CDM models, along with their collider signatures, have been discussed in,
{\em e.g.},  Refs.\ \cite{1309.2986,1310.7901,1406.0617,1612.08621,1801.09115,1808.03352}.
\end{enumerate}

It must be admitted at this point that these models, {\em per se}, fall short of explaining the CMS diphoton signal at
96 GeV. While the models include such a scalar, its diphoton decay width is either zero or too small to
be interesting. This is because $\chi$ does not couple directly to photons (for decay) or gluons (for production), 
neither does it couple to the SM quarks and leptons through which production and decay can be 
mediated\footnote{Unless $\chi$ has a significant doublet component, which, however, is ruled out from the 
LHC data.}. To alleviate this shortcoming, we invoke either of these 
two further extensions to each of the three models, which are:

Subclass (a): The model contains an $SU(2)_{\rm L}$ singlet vectorial quark $Q$. This is in addition to the 
vectorial fermion in the RSF or CSF models, and may not belong to the same multiplet. 
To avoid direct detection at the LHC, we take, for our analysis, $M_Q = 2$ TeV. The relevant part of the 
Lagrangian, including both QED and QCD interactions, looks like
\be
{\cal L} = \overline{Q}\left( i\gamma^\mu\partial_\mu - m_Q\right) Q 
- qe\, \overline{Q}\gamma^\mu Q A_\mu - g_s\, \overline{Q}_i \gamma^\mu (T^a)_{ij} Q_j G^a_\mu\,,
\label{eq:VQLag}
\ee
and once the Yukawa couplings are introduced, the physical mass $M_Q$ may contain, apart from $m_Q$, 
a contribution from symmetry breaking. 
As this is a vectorial 
fermion, no gauge or mixed anomaly is introduced, neither do we encounter the risk of a large nonperturbative 
Yukawa coupling (with the singlet scalar $\chi$) as most of the mass can originate without symmetry breaking. 
In our subsequent discussion, we will take its Yukawa coupling to be $\sim \frac23$ in the RSF model 
and $\sim 0.95$ in the CS and CSF models, and its electric charge to be
$+2e$. This correctly reproduces the number of diphoton events $\sim 370$ in the $\sqrt{s}=8$ TeV 
data. It could have been 
done with a smaller charge and a large Yukawa coupling too, but that makes the potential of the theory 
unstable even at the LHC scale. We will discuss these issues later. 

Subclass (b): We do not invoke any extra fermion or any extra degrees of freedom; rather, in the spirit of 
effective theories, we integrate out the new fields and introduce two dimension-5 operators of the form 
$F_{\mu\nu}F^{\mu\nu}\chi$ and $G^a_{\mu\nu}G^{a\mu\nu}\chi$, for the photon and the gluon fields 
respectively, where the first term is responsible for the 
diphoton decay and the second term for the production of $\chi$ through gluon-gluon fusion. This is obviously not
an independent subclass, as the heavy fermions of Subclass (a) can be integrated out to generate Subclass (b). 
On the other hand, many other extensions can also lead to such effective operators. 
We use the effective Lagrangian
\be
{\cal L}_{\rm eff} = -\frac{C_\gamma}{\Lambda}\, F_{\mu\nu}F^{\mu\nu}\chi - \frac{C_g}{\Lambda}\, 
G^a_{\mu\nu} G^{a\mu\nu}\chi\,,
\label{eq:dim-5}
\ee
and the Wilson coefficients (WC) $C_\gamma$ and $C_g$ reproduce the CMS diphoton rate for 
moderate enough values even with $\Lambda = 100$ TeV, say $C_\gamma \sim 0.30$ and 
$C_g \sim 0.54$. 
One can even have higher-dimensional operators, like $F_{\mu\nu}F^{\mu\nu}S^2$, but their WCs will 
be suppressed by higher powers of $\Lambda$. 

One may note here that such models, either with a vectorlike fermion or effective operators, were
discussed in detail during the heydays of the now-dead 750 GeV diphoton resonance. We refer
the reader to, {\em e.g.}, Refs.\ \cite{750gev} for a detailed analysis, and to Refs.\ \cite{750dark} 
for its association with possible dark matter models.

We will concentrate on the allowed parameter space for all these models, taking the theoretical and experimental 
constraints, including the CMS data, into account. 
One major constraint is the stability of the potential; when the couplings evolve with 
energy, the potential should neither be unbounded from below, nor should the couplings blow up at the 
Landau pole (a more conservative statement is that they should remain perturbative). Up to what scale the 
potential should be well-behaved? Ideally, it is the Planck scale $M_{\rm Pl}$, but that makes the parameter space too 
restrictive; even the SM may not be stable up to $M_{\rm Pl}$. In fact, this scale can be anything beyond the 
reach of the LHC, where some new degrees of freedom appear and cure the bad behaviour of the potential. 
For Subclass (b), we demand that the theory be well-behaved up to the scale $\Lambda$. 
We take this scale to be 100 TeV for all our subsequent discussion. As we will see, the scalar couplings 
of the potential get bounded both from above and below with the conditions of stability and triviality. 
Pushing up the scale further only squeezes the parameter space, but no qualitative change takes place.

The paper is organised as follows. In Section \ref{sec:models}, 
we give a brief outline of these models, including the theoretical and experimental constraints.
The final parameter space for all
the three models, taking into account the aforementioned constraints plus the direct CDM search results,
are shown in Section \ref{sec:results}, where we also enlist a 
few novel observations. Section \ref{sec:conclude} summarises and concludes the paper.

\section{The Models} \label{sec:models}

As mentioned in the Introduction, we will discuss three models in this Section. The common characteristics of
these three models are: (i) they have one (or more) potential CDM candidate(s), and possibly some other degrees 
of freedom to generate the CMS diphoton signal; (ii) apart from the 125 GeV
scalar resonance $h$, which is dominantly the SM doublet Higgs boson, every model is constructed in such a way
as to further accommodate another scalar $\chi$ at 96 GeV. Even if the 96 GeV bump does not stand the test of
time, the results, on the allowed parameter space of the CDM mass and couplings, will still be more or less
valid, although some constraints will be relaxed. 

The following constraints on the potential need to be considered:
\begin{itemize}

\item Existence of a definite ground state, which
means that the potential cannot be unbounded from below along any direction in the field space. As 
mentioned before, we allow the possibility of some new physics to take over at a scale $\Lambda$, 
so all we need is a well-defined ground state of the potential up to $\Lambda$, beyond which the new physics
may cure any possible malady and make the potential well-behaved. We take $\Lambda = 100$ TeV.  

\item Partial wave unitarity, which essentially leads to the conservation of probability.
For any scattering we can decompose the amplitude into partial waves
\be
A = 16\pi \sum_{\ell=0}^\infty \, (2\ell+1) \, P_\ell(\cos\theta)\, a_\ell,
\ee
and by virtue of the optical theorem which relates the cross-section with the imaginary part 
of the amplitude for zero scattering angle, one gets \footnote{To treat all possible spins of incoming 
and outgoing particles, one should use Wigner's $D$-functions, but $D^0_{00}$ is directly
related with $a_0$.}
\be
|a_\ell|^2 = {\rm Im} \, a_\ell \, \, \Rightarrow\, \,  {\rm Re}\, a_\ell \leq \frac12\,.
\ee
We will be interested in $\ell=0$ partial waves only. The best bounds are obtained when one diagonalises 
the $S$-matrix and uses the scattering from one eigenvector channel $|{\rm in}\ket$ to another $|{\rm
out}\ket$. They may be the same eigenvectors. We follow Ref.\ \cite{goodsell} to get the unitarity 
constraints.

\item Triviality, or the constraint that none of the 
couplings hit the Landau pole below $\Lambda = 100$ TeV. As we will see later, these two conditions 
turn out to be almost equivalent. Stability and triviality, taken together, limit the range of the scalar parameters of the 
potential.

\item The invisible decay width of the Higgs when light
scalars and/or fermions are present in the model should be less than 19\%  \cite{1905.07150}. This is 
relevant only if the CDM mass is less than $m_h/2$. 

\item The dominantly doublet nature of $h$, as established by the LHC data, which means that the mixing 
angle with the singlet scalar must be small. This necessitates the introduction of either new fermions or new
effective operators to explain the $\chi\to \gamma\gamma$ decay rate.  

\item The constraints coming from the CDM direct detection experiments, interpreted in terms of a thermalised 
dark matter, as well as those coming from the relic density of the CDM, {\em i.e.}, the universe must not be
overclosed \cite{1608.07648,Cui:2017nnn,Aprile:2017iyp}. 
The bounds from LUX, XENON1T, and PandaX-II 
collaborations are quite close to each other in the CDM mass region that we are interested in. The 2018 update 
of XENON1T gives slightly stronger bounds \cite{Aprile:2018dbl}, but does not affect our results in a significant 
way if $m_{\rm CDM} > m_\chi/2$. 
For $m_{\rm CDM} > m_\chi$, the CDM pair annihilation to $\chi\chi$ 
controls $\Omega_{\rm CDM}$. For $m_b < m_{\rm CDM} < m_\chi$, the CDM pair annihilation rates to 
$b\bar{b}$, diphoton and digluon are of the same order.

\item Bounds on the oblique electroweak parameters, which, however, have a completely negligible effect 
if the singlet-doublet mixing is tiny, and/or the new fermion is a gauge singlet with no mixing to the SM 
fermions. This is true even for the new charged fermions in Subclass (a).
\end{itemize}

The parameter space is constrained assuming the CDM candidate(s) 
being thermalised and satisfying the relic density limit, {\em i.e.},
there are no other particles that contribute to the relic density. To compute the relic density one solves 
the Boltzmann equation \cite{1412.1105} for $Y=n/s$, where $n$ is the number density and $s$ is the 
entropy density, and calculates the thermal averaged cross-section. Assuming a scalar CDM $S$ of mass $m_S$
and a coupling of the form $(\delta_2/4) \Phi^\dag\Phi S^2$, the spin-independent CDM-nucleon
cross-section is given by
\be
\sigma_{SI}= \frac{\delta_2^2 m_N^4 f^2}{16 \pi m_{S}^2 m_h^4}\,,
\ee
where $m_N$ is the nucleon mass, and $f$ is the nuclear form factor, often taken to be $\sim \frac13$ 
\cite{strumia}. In this paper, we take the CDM to be thermalised, and find those regions of the parameter space 
that reproduce the correct CDM density. 
There can also be loop-induced CDM-nucleus scattering diagrams \cite{1803.05660}. However, the loop
suppression makes them negligible compared to the tree-level amplitudes. If the CDM is a scalar, the loop
amplitude is a few orders of magnitude smaller than the tree-level one. For a fermionic CDM $\psi$, the 
Higgs-$\psi$ coupling is suppressed by the scalar mixing angle, but even then the loop suppression
works in favour of the tree-level amplitudes for the mass ranges that we are interested in.

Before we go into the three models, let us have a brief recapitulation of the Real Singlet (RS) model, which
is SM plus a real gauge singlet scalar $S$. The scalar potential can be written as
\be
V(\Phi, S)_{\rm RS} = -\frac{m^2}{2} \Phi^{\dag} \Phi +\frac{b_2}{4} S^2 +\frac{\lambda}{4} \left(\Phi^{\dag}\Phi\right)^2
+\frac{\delta_2}{4} \Phi^{\dag} \Phi  S^2 +\frac{d_2}{16} S^4\,,
\label{eq:V-RS}
\ee
where $S$-odd terms are banished from the potential by a $\mzt$ symmetry: $S\to -S$. 
We will denote the CP-even neutral component of $\Phi$ by $\phi$, with 
\be
\bra \phi \ket \equiv v= 246~{\rm GeV}\,.
\ee 
In the absence of $\Phi$-$S$ mixing, $\phi=h$, where $h$ is the physical 125 GeV scalar resonance. 

If $S$ does not get any vacuum expectation value (VEV), it can act as a potential CDM candidate. 
However, as has been shown in Refs.\ \cite{1412.1105,1705.07931,1710.08723,1806.11281}, the allowed 
region for $m_S < m_h$ is rather fine-tuned; the CDM solution survives only in a narrow resonance region
around $m_S \approx m_h/2$. There are other allowed regions for large values of $m_S$, like $m_S > 1$ TeV, 
where $S$ can be a viable CDM candidate. If we allow $S$ to be a component of CDM, and not the 
only constituent, the allowed region starts from $m_h \approx m_S$. All these situations have been extensively
explored. We will not discuss this model any further, but we would like to have a digression here on the 
CMS signal, which is applicable to all the other models that we discuss.

\subsection{Digression: The CMS diphoton signal}

If $S$ mixes with the SM doublet $\Phi$, it cannot be a CDM candidate, but can the lighter mass eigenstate, 
which is dominantly a singlet, act as $\chi$? The answer, unfortunately, is no; the mixing must be small to keep 
the doublet nature of $h$ consistent with the LHC data, and that makes both the production cross-section, 
as well as the diphoton decay width, so small as to be completely unobservable at the LHC. We may refer the 
reader to Ref.\ \cite{1906.02175} for further discussion.

Introduction of another vector singlet quark $Q$ with an electric charge $+q$ solves the problem. Apart from
Eq.\ (\ref{eq:VQLag}), we add a Yukawa term to the Lagrangian:
\be
{\cal L}_1 = - h_Q \, \overline{Q} Q S\,.
\ee
The production cross-section for $gg\to \chi$ is given by 
\be
\sigma(gg\to\chi) = \frac{\alpha_{s}^{2}}{576 \, \pi} \, h_{Q}^{2} \, \cos^{2}\alpha \, \frac{m_{\chi}^{2}}{M_{Q}^{2}} \, \frac{d{\cal L}^{gg}}{d m_{\chi}^{2}}\,,
\ee
where $d{\cal L}^{gg}/d m_{\chi}^{2}$ is the gluon luminosity using the MSTW2008NLO parton distribution
\cite{stirling}, and $\alpha$ is the singlet-doublet mixing angle\footnote{For all
practical purpose, $\cos\alpha \approx 1$.}.
The decay width of $\chi\to\gamma\gamma$ is 
\be
\Gamma(\chi\rightarrow\gamma\gamma)=\frac{N_{c}^{2} \alpha_{\rm em}^{2} q^{4}}{72 \pi^{3}} \, 
h_{Q}^{2}  \cos^{2}\alpha \, \frac{m_{\chi}^{3}}{M_{Q}^{2}}\,. 
\ee
To generate the required number of events, we take $h_Q \approx \frac23$ and $|q|=2$, with which one has 
about 370 (1070) events for $\sqrt{s}=8\, (13)$ TeV. 
If there are two singlets, as in the CS and CSF models, the Yukawa interaction is of the form 
$(h_Q/\sqrt{2})\, \bar{Q}Q(S_1+iS_2)$, and so we need to scale up $h_Q$ by $\sqrt{2}$ to $h_Q\approx 0.95$.
With lower values of $|q|$, one needs higher $h_Q$, but this 
makes the potential unstable before 100 TeV, through the renormalisation group evolution of the singlet
quartic coupling $d_2$, with a contribution going as $-h_Q^4$.

Alternatively, one may introduce two dimension-5 operators, as shown in Eq.\ (\ref{eq:dim-5}). Two representative 
benchmark values for $(C_{\gamma}, C_{g})$ to generate required number of events may be taken as 
$(0.30, 0.54)$ and $(0.28, 0.34)$ for $\sqrt{s}=8$ TeV and $\sqrt{s}=13$ TeV respectively. 

We will now briefly discuss our models, with the implicit understanding that such a mechanism to generate 
the required number of $\chi\to\gamma\gamma$ decays is added to all of them. One important point to note right
here is that the CDM mass must have a lower bound of approximately $m_\chi/2 \approx 48$ GeV. For a lighter
CDM, $\chi$ dominantly decays to a pair of CDM, which suppresses the diphoton branching ratio by a few 
orders of magnitude.

\subsection{The Complex Scalar Model (CS)}

Extending the scalar sector by a complex singlet $S\equiv (S_1 + i S_2)/\sqrt{2}$, 
the most general renormalisable potential is of the form \cite{barger-complex}

\bea
V(\Phi,S)&=& -\frac{m^2}{2} \Phi^{\dag}\Phi + 
\frac{\lambda}{4}\left( \Phi^{\dag}\Phi\right)^2 + 
\left( \frac{\delta_1}{4}e^{i\theta_{\delta_1}} \Phi^{\dag} \Phi S + {\rm c.c}\right) + 
 \frac{\delta_2}{2}\Phi^{\dag}\Phi \mid S \mid^2 +
\n
\\
&& \left( \frac{\delta_3}{4}e^{i\theta_{\delta_3} }\Phi^{\dag} \Phi S^2+{\rm h.c}\right ) + 
\left( a_1 e^{i\theta_{a_1}} S+ {\rm h.c}\right) + 
\left( \frac{b_1 e^{i\theta_{b_1}}}{4} S^2 + {\rm h.c}\right)+
\n
\\
&& \frac{b_2}{2}\mid S \mid^2 +
\left( \frac{c_1 e^{i\theta_{c_1}}}{6}S^3+ {\rm h.c}\right) +
\left( \frac{c_2 e^{i\theta_{c_2}}}{6}S\mid S \mid^2 + {\rm h.c}\right) +
\n \\
&& \left( \frac{d_3 e^{i\Phi_{d_3}}}{8} S^2 \mid S \mid^2+ {\rm h.c}\right) 
+ \frac{d_2}{4}\mid S \mid^4\,,
\label{eq:CSpot}
\end{eqnarray}
where the couplings are taken to be real apart from an explicit phase factor, generically written as 
$\exp(i\theta)$, with $0\leq\theta\leq \pi$. Apart from this, one should include the Yukawa coupling with $Q$:
\be
{\cal L}_1 = -\frac{h_Q}{\sqrt{2}} \, \overline{Q} \left(S_1 + i\gamma_5 S_2\right) Q\,,
\label{eq:VQLag2}
\ee
or suitable effective operators for Subclass (b). We will first try to
confine ourselves to a simpler case, by reducing the number of independent parameters through a global
$U(1)$ symmetry on $S$: $S\to S\exp(i\zeta)$. This simplifies the potential to
\be
V(\Phi,S)=-\frac{m^2}{2}\Phi^{\dag} \Phi + 
\frac{\lambda}{4}\left(\Phi^{\dag}\Phi\right)^2 + 
\frac{\delta_2}{2} \Phi^{\dag} \Phi \mid S \mid ^2 + 
\frac{b_2}{2}\mid S \mid ^2 +\frac{d_2}{4}\mid S \mid ^4\,,
\ee
but leads to a massless Goldstone boson in the spectrum when the symmetry breaks spontaneously to give
$S$ a VEV. To avoid this, we introduce a soft $U(1)$ breaking term (but still keeping the $\mzt$ of $S\to -S$ 
intact) in the potential:
\be
V_{\rm CS}(\Phi,S) =  -\frac{m^2}{2}\Phi^{\dag} \Phi + 
\frac{\lambda}{4}\left(\Phi^{\dag}\Phi\right)^2 + 
\frac{\delta_2}{2} \Phi^{\dag} \Phi \mid S \mid ^2 + 
\frac{b_2}{2}\mid S \mid ^2 +\frac{d_2}{4}\mid S \mid ^4
+\left( \frac{b_1}{4}e^{i\theta}S^2 + {\rm h.c.}\right)\,,
\label{eq:CSpot2}
\end{equation}
so that $S_2$ becomes a pseudo Nambu-Goldstone boson with a mass proportional to $b_1$.
The phase $\theta$ helps in realising the vacuum alignment condition. We take $\theta = \pi$ and let $S_1$ get 
the VEV:
\be
\bra S \ket = \frac{1}{\sqrt{2}} \bra S_1 \ket = s_1/\sqrt{2}\,.
\ee
In terms of the component fields, the potential is
\be
V_{\rm CS}(\phi,S_1,S_2) = 
-\frac{m^2}{4}\phi^2+\frac{\lambda}{16}\phi^4 +
\frac{\delta_2}{8} \phi^2\left(S_1^2+S_2^2\right)+\frac{b_2}{4}\left(S_1^2+S_2^2\right) +
\frac{d_2}{16}\left(S_1^2+S_2^2\right)^2-\frac{b_1}{4}\left(S_1^2-S_2^2\right)\,,
\label{eq:CSpot3}
\ee
with the extremisation conditions
\bea
\frac{\partial V}{\partial \phi} &=&
-\frac{m^2}{2}\phi + \frac{\lambda \phi^3}{4}+\frac{\delta_2}{4} \phi (S_1^2+S_2^2) = 0\,,\n\\
\frac{\partial V}{\partial S_1} &=&
\frac{S_1}{2}\left( b_2-b_1+\frac{\delta_2 \phi^2}{2}+\frac{d_2}{2}(S_1^2+S_2^2)\right) = 0\,,\n\\
\frac{\partial V}{\partial S_2} &=&
\frac{S_2}{2}\left( b_2+ b_1 +\frac{\delta_2}{2}\phi^2+\frac{d_2}{2}(S_1^2+S_2^2)\right)= 0\,.
\eea
Note that the terms of the RS and the CS potentials have been written in such a way as to ensure 
identical CDM couplings with the Higgs as well as its self-coupling. 

Obviously, the solution with $\bra\phi\ket = v = 0$ is not physical. For $v\not= 0$, there can be two possible 
cases that will give rise to a CDM candidate: (i) $s_1 = s_2 = 0$, and $s_1 \not=0$, $s_2 = 0$. In the first
case, there is no mixing between $\phi$ and $S_{1,2}$, and hence no $\chi$, so we will drop that from our focus.
Only the second case is interesting, so let us treat that in more detail.
 
The conditions $v\not= 0$, $s_1\not= 0$, $s_2 = 0$ lead to
\bea
s_1^2 &=& \frac{2\lambda\left( b_2-b_1\right) + 2\delta_2 m^2}{\delta_2^2-\lambda d_2}\,,\n\\
v^2 &=& \frac{2\delta_2 \left( b_2- b_1 \right) + 2d_2 m^2}{\lambda d_2-\delta_2^2}\,,
\eea
thus ensuring that the four minima at $(\pm \sqrt{v^2}, \pm \sqrt{s_1^2})$ are of equal depth, which is a consequence 
of the still-intact $\mzt$ symmetry. 

The couplings can be further constrained from the existence of a well-defined ground state: 
\be
\lambda > 0\,, \ \ \ d_2 > 0\,,\ \ \  \lambda d_2 > \delta_2^2\,.
\ee

The mass matrix for the neutral scalars is
\be
M^2 = \frac12
\begin{pmatrix}
\lambda v^2 & \delta_2 v s_1 & 0 \cr
\delta_2v s_1 & d_2 s_1^2 & 0 \cr
0 & 0 & 2b_1
\end{pmatrix}\,.
\ee
with the mass eigenvalues
\bea
m^2_{h,\chi} &=& 
\frac{\lambda v^2}{4}+\frac{d_2 s_1^2}{4} \pm \sqrt{\left( \frac{\lambda v^2}{4}-\frac{d_2 s_1^2}{4}\right)^2
+\frac{\delta_2^2 v^2 s_1^2}{4}}\,,\n\\ 
m^2_{S_2} &=& b_1\,.
\eea
$S_2$ is a stationary state itself; there is no mixing with the other scalars as $s_2 = 0$. 

The $h$-$\chi$ mixing angle $\alpha$ is given by
\be
\tan{(2\alpha)}=\frac{2\delta_2 vs_1} {\lambda v^2 - d_2 s_1^2 }\,.
\ee

For the analysis, it is helpful to express the parameters of the potential, namely,
$\lambda$, $d_2$, $\delta_2$, $m$, $b_2$, and  $b_1$, in terms of the VEVs $v$ and $s_1$, the 
mixing angle $\alpha$, and the three masses $m_h$, $m_\chi$ and $m_{S_2}$. Among these, $m_h$, 
$m_\chi$, and $v$ are known, so there are only three free parameters, including $\alpha$, 
which is known to be small. 

The transformation equations between the two bases are

\bea
\lambda &=& \frac{2}{v^2} \left( m_h^2 \cos^2\alpha + m_\chi^2\sin^2\alpha\right)\,,\n\\
d_2    &=& \frac{2}{s_1^2} \left( m_h^2 \sin^2\alpha + m_\chi^2\cos^2\alpha\right)\,,\n\\
\delta_2 &=& \frac{\left(m_{h}^2-m_\chi^2\right)\, \sin(2\alpha)}{v s_1}\,,\n\\
m^2 &=& \frac12 \left(\delta_2 s_1^2+ \lambda v^2\right)\,,\n\\
b_1-b_2 &=& \frac12 \left(d_2 s_1^2 + \delta_2 v^2\right)\,,\n\\
b_1 &=& m_{S_2}^2\,.
\label{eq:reln1}
\eea

We would also like to see how the dimensionless couplings evolve with energy. 
For this, we incorporate the explicit introduction of the vectorial quark $Q$, {\em i.e.}, 
Subclass (a), and will do the same thing for the other two models. The one-loop $\beta$-functions are
\cite{machacek-vaughn}
\bea
16\pi^2 \beta_{d_2} &=& 5 d_2^2 + 2 \delta_2^2 + 12 d_2 h_Q^2 -24 h_Q^4\,,\n\\
16 \pi^2 \delta_2 &=&
\delta_2 \left(2 d_2 + 2 \delta_2 - \frac{3}{2} g_1^2 - \frac{9}{2} g_2^2 + 6 h_Q^2 + 3 \lambda +6 g_t^2\right)\,,\n\\
16 \pi^2 \beta_{\lambda} &=& 
\left( \delta_2^2 + \frac{3}{2} g_1^4 + 3 g_1^2 g_2^2 + 
\frac{9}{2}g_2^4 - 3 g_1^2 \lambda - 9 g_2^2\lambda +  6 \lambda^2 + 12\lambda g_t^2-24 g_t^4\right)\\
16\pi^2\beta_{h_Q} &=& 4 h_Q^3 - 8 h_Q\, \left(3 g_1^2 + g_3^2\right) \,.
\label{CSrge}
\eea
with $\beta_x = dx/dt$, and $t\equiv \ln(q/q_0)$, where $q_0$ is some reference scale 
to set the boundary conditions on the couplings, and $q$ is the relevant energy scale where they are measured.
Higher orders do not affect our results in any appreciable way.
We must ensure that nowhere in the parameter space the triviality bound is reached below 100 TeV. 

The parameter space becomes more complicated with the introduction of several new couplings if we allow the 
breaking of the $\mzt$ symmetry\footnote{
Even the Yukawa couplings break the $\mzt$ symmetry of the potential through one-loop diagrams.
However, that results in an explicit breaking, not spontaneous, so the issue of domain walls between different 
vacua does not arise.}. 
However, qualitatively it does not add much over the parameter space 
of the CS model that we will discuss in the next Section. One important modification is that the vacua are 
no longer degenerate, and one has to put the additional constraint that either the SM vacuum is the deepest
one, or it is a false vacuum with the tunnelling time larger than the age of the universe. We will not discuss
the $\mzt$ breaking case any further\footnote{The effective operators of the type $F^2S$ and $G^2S$ break the 
$\mzt$ symmetry.}.

\subsection{Real singlet scalar with vectorial fermion (RSF)}

Let us add one vector singlet fermion $\psi$ to the real singlet model of Eq.\ (\ref{eq:V-RS}). 
This leads to the generalised potential\footnote{The last two terms do not belong to the scalar potential,
but the generalised potential includes all terms that are not kinetic.} 
\be 
V(\Phi,S,\psi)= -\frac{m^2}{2}\Phi^{\dag}\Phi + \frac{\lambda}{4}\left(\Phi^{\dag}\Phi\right)^2
+\frac{\delta_2}{4} \Phi^{\dag} \Phi S^2+\frac{b_2}{4} S^2 +\frac{d_2}{16} S^4
+ m_{\psi}\bar{\psi}\psi + (Y/\sqrt{2})\, \bar{\psi}\psi S\,,
\label{eq:V-RSF}
\ee
as well as the fermionic 
with $Y$ being the Yukawa coupling, whose presence already breaks the ${\mathbb {Z}}_2$ symmetry of 
$S \to -S$. The CDM candidate, $\psi$, is stable as long as it does not mix with the neutrinos. For Subclass (a), 
one should also add the Yukawa term in Eq.\ (\ref{eq:VQLag}).

Using the same notation as for the CS model, we get the scalar mass eigenstates as
\be
m^2_{h,\chi} = 
\frac14\left[ 
\lambda v^2 + d_2 s_1^2 \pm \sqrt{\left( \lambda v^2-d_2 s_1^2 \right)^2
+4 \delta_2^2 v^2 s_1^2} \right]\,,
\label{eq:rsf-1}
\ee
with the mixing angle
\be
\tan{(2\alpha)}=\frac{2\delta_2 vs_1} {\lambda v^2 - d_2 s_1^2 }\,.
\label{eq:rsf-2}
\ee
The fermion $\psi$ is the CDM candidate, whose mass is $m_{\rm CDM} = m_\psi + Y s_1/\sqrt{2}$. 

\begin{figure}[h]
\centerline{ \includegraphics[origin=c,width=9cm]{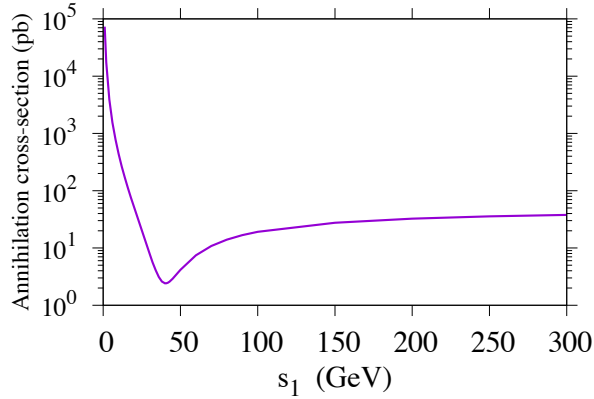}
}
\vspace{-0.8cm}
\caption{\small{The dark matter annihilation cross-section as a function of the singlet scalar VEV, $s_1$,
in the RSF model. The cross-section dips at $s_1\sim 50$ GeV, due to destructive interference of 
the $s$- and $t$-channel amplitudes. For more details, see text. The typical annihilation cross-section 
for WIMP-like CDM is ${\cal O}(10^{-8})$ GeV$^{-2}$ or a few pb, although it depends on the mass of the CDM
and its freeze-out temperature. 
}}
\label{fig:cdm-rsf0}
\end{figure}

The inclusion of $\psi$ helps in stabilising the potential: The negative contribution in the 
$\beta$-function keeps $\delta_2$ and $d_2$ in check, which in turn affects the running of $\lambda$. 
The one loop $\beta$-functions are given by
\bea
16 \pi^2 \beta_{d_2} &=& \frac92 d_2^2 + 2 \delta_2^2 + 24 d_2 h_Q^2 - 96 h_Q^4+ 4 d_2 Y^2 - 8 Y^4\,,\n\\
16 \pi^2 \beta_{\delta_2} &=&
  \delta_2 \left( \frac32 d_2 + 2 \delta_2 - \frac{3}{2} g_1^2 - \frac{9}{2} g_2^2 + 12h_Q^2 + 
     2 Y^2 + 3 \lambda  + 6g_t^2\right)\,,\n\\
16 \pi^2 \beta_{\lambda} &=& \frac{\delta_2^2}{2} + 
\frac{3}{2}g_1^4 + 3 g_1^2 g_2^2 + 
\frac{9}{2}g_2^4 - \left(3 g_1^2 + 9 g_2^2\right) \lambda + 6\lambda^2 + 
12\lambda g_t^2 -  24 g_t^4\,,\n\\
16\pi^2\beta_{h_Q} &=&  9 h_Q^3 + h_Q Y^2 - 8h_Q\, \left( 3 g_1^2 + g_3^2\right) \,,\n\\
16\pi^2\beta_{Y} &=& Y\, \left( 6h_Q^2+\frac52 Y^2\right)\,.
\eea
With more scalars added to the SM, the scalar quartic couplings
tend to blow up, hitting the Landau pole much below the Planck scale. One has to choose points in the 
parameter space such that not only Eq.\ (\ref{eq:omegadm}) is satisfied, but the couplings also 
remain perturbative up to $\Lambda= 100$ TeV. 
As we will see later, this puts some nontrivial constraints on the parameter space. 

In Fig.\ \ref{fig:cdm-rsf0}, we show the annihilation cross-section of the CDM, $\psi\bar\psi \to X$, 
where $X$ is any allowed SM particle-antiparticle pair, or even $\chi\chi$. 
Note that $\psi$ is non-relativistic even at freeze-out and hence its number density goes as
$T^{3/2}\exp(-m_\psi/T)$, where $T$ is the temperature of the CDM.
The $s$-channel amplitude is
mediated by the singlet scalar $S$ (and hence by $h$ and $\chi$ after mixing), while the $t$-channel 
process is simply $\psi\bar\psi \to SS$. The amplitudes depend on the Yukawa coupling $Y$. For this plot, 
the CDM mass is kept fixed at 100 GeV, so the cross-section, in turn, depends on the VEV $s_1$. 
We find that there is a dip in the cross-section near $s_1\approx 50$ 
GeV, as the $s$- and $t$-channel amplitudes interfere destructively. The position of the dip, obviously, will 
depend on the chosen value of $m_\psi$ and $Y$. This fact will be relevant when we study the allowed parameter
space.

\subsection{Complex singlet scalar with vectorial fermion (CSF)}

Analogous to RSF, the particle content is that of the CS model plus the singlet vectorial 
fermion $\psi$. The generalised potential looks like
\be
V_{\rm CSF}(\Phi,S,\psi) = V_{\rm CS} + m_\psi \bar{\psi}\psi + \frac{Y}{\sqrt{2}}\, \left(\bar{\psi}\psi S_1 + 
i\bar\psi \gamma_5 \psi S_2\right)\,,
\ee
augmented by the Yukawa term in Eq.\ (\ref{eq:VQLag2}), where $V_{\rm CS}$ is given by Eq.\ (\ref{eq:CSpot2}). 
This can lead to two distinct possibilities:\\
(i) $\bra \Phi\ket = v/\sqrt{2}$, $\bra S_1\ket = s_1$, $\bra S_2\ket = 0$: The scalar sector will be 
completely identical to that of the CS model, and there are two possible CDM candidates, namely, $\psi$ and $S_2$. 
Based on the number of CDM candidates, we will call this model CSF-2. \\
(ii) $\bra S_2\ket = s_2\not= 0$. This leads to $\phi$-$S_1$-$S_2$ mixing. In other words, the field $S$ gets a complex
VEV $s=s_1+is_2$. The only CDM candidate is $\psi$. This will, therefore, be called the CSF-1 model
\footnote{In both CSF-1 and CSF-2, both the singlet scalars have a fermion loop introduced contribution
to their respective masses.}

The phenomenology of CSF-1 is more or less the same as that of RSF, so we will concentrate on the CSF-2
model. The scalar mass eigenstates are given by Eq.\ (\ref{eq:rsf-1}), as well as 
$m^2_{S_2} = b_1$, while the mass of $\psi$ is, as before, $m_{\rm CDM}= m_{\psi}+Y s_1/\sqrt{2}$.
The $h$-$\chi$ mixing angle $\alpha$ is the same as shown in Eq.\ (\ref{eq:rsf-2}). 
The one loop $\beta$-functions are:
\bea
16 \pi^2 \beta_{d_2} &=& 5 d_2^2 + 2 \delta_2^2 + 12 d_2 h_Q^2 -24 h_Q^4 + 4 d_2 Y^2 - 8 Y^4\,,\n\\
16 \pi^2 \beta_{\delta_2} &=&
  \delta_2\, \left(2 d_2 + 2 \delta_2 - \frac{3}{2} g_1^2 - \frac{9}{2} g_2^2 + 6 h_Q^2 + 
     2 Y^2 + 3 \lambda + 6 g_t^2\right)\,,\n\\
16\pi^2\beta_{\lambda} &=&
  \delta_2^2 + \frac{3}{2}g_1^4 + 3 g_1^2 g_2^2 + \frac{9}{2}g_2^4 - \left( 3 g_1^2 + 9 g_2^2\right) \lambda 
  + 6 \lambda^2 + 12\lambda g_t^2-24 g_t^4\,,\n\\ 
16 \pi^2\beta_{h_Q} &=& 4 h_Q^3 + h_Q Y^2  - 8h_Q\, \left(3 g_1^2 + g_3^2\right) \,,\n\\ 
16 \pi^2\beta_{Y} &=& Y \left(3 h_Q^2+2 Y^2\right)\,.
\eea

\begin{figure}[h]
\includegraphics[width=5.5cm]{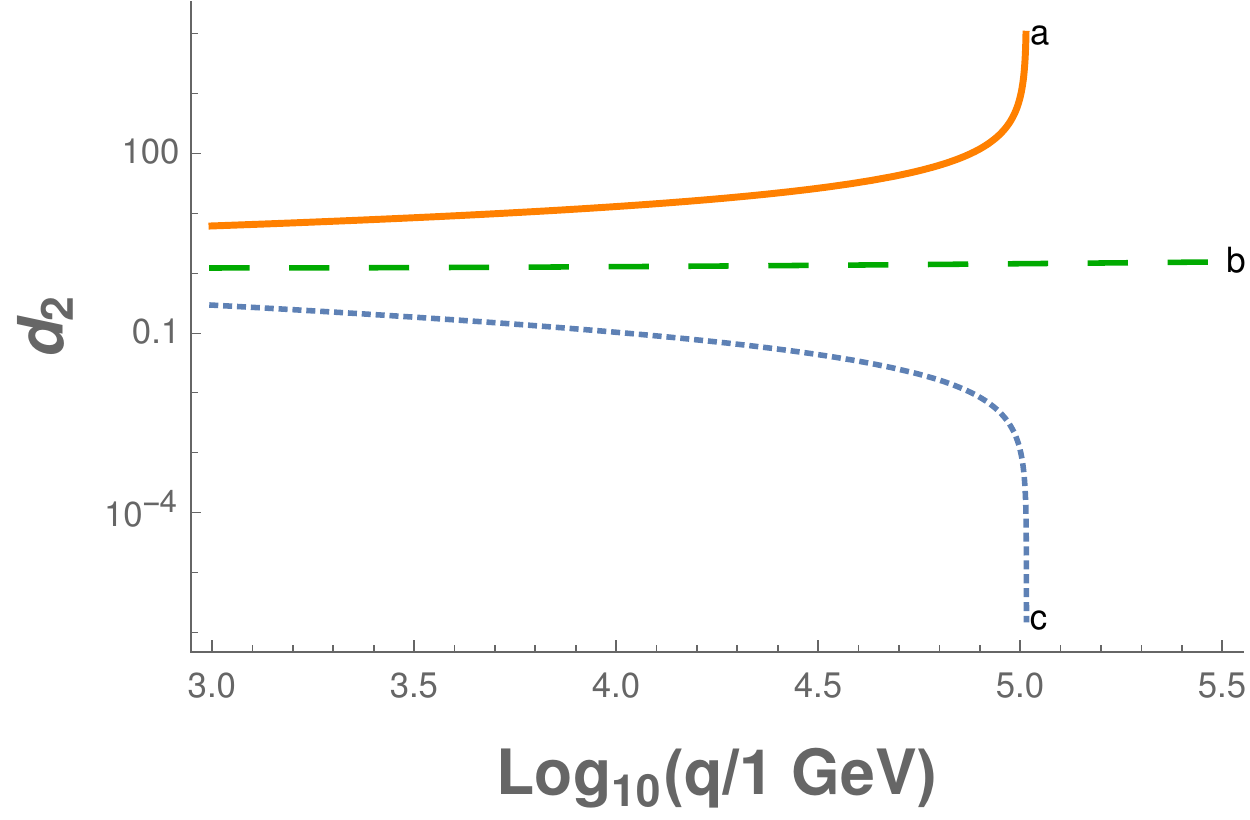} \ \ 
\includegraphics[width=5.5cm]{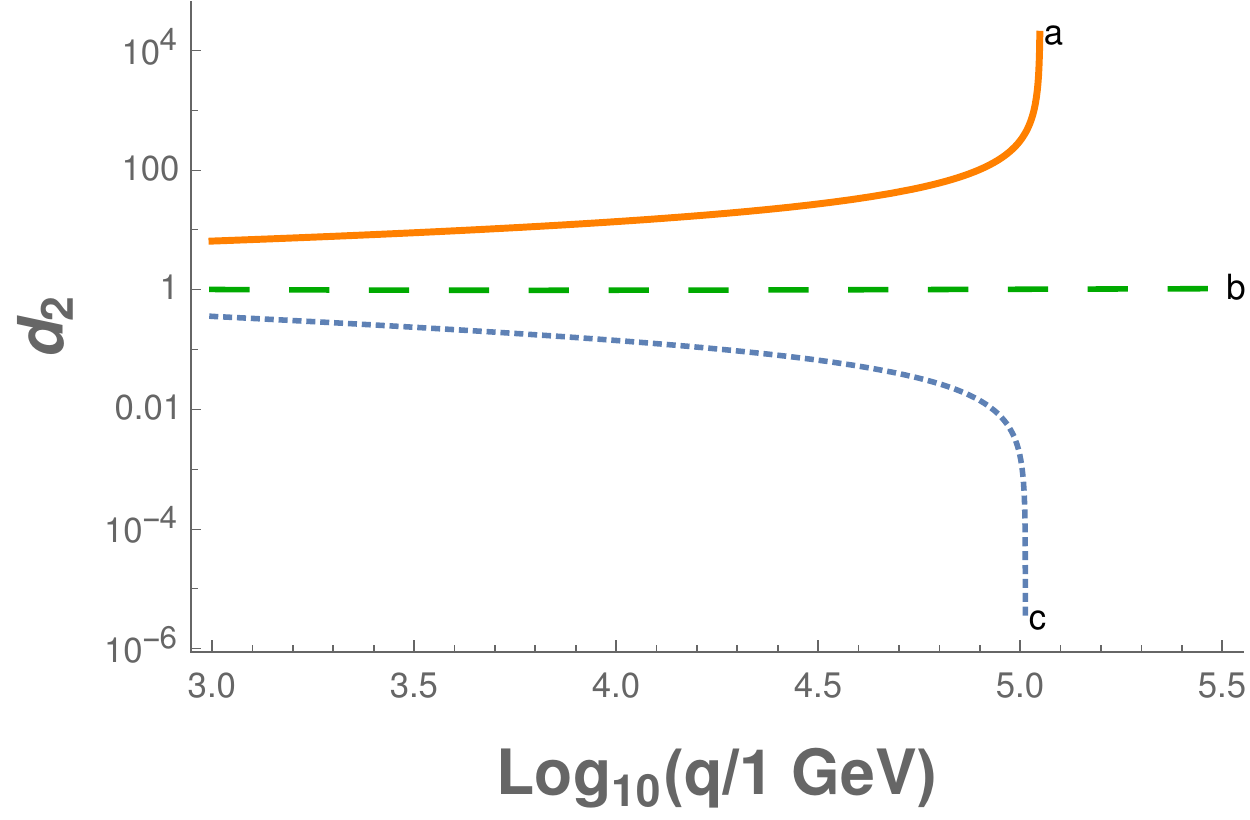} \ \ 
\includegraphics[width=5.5cm]{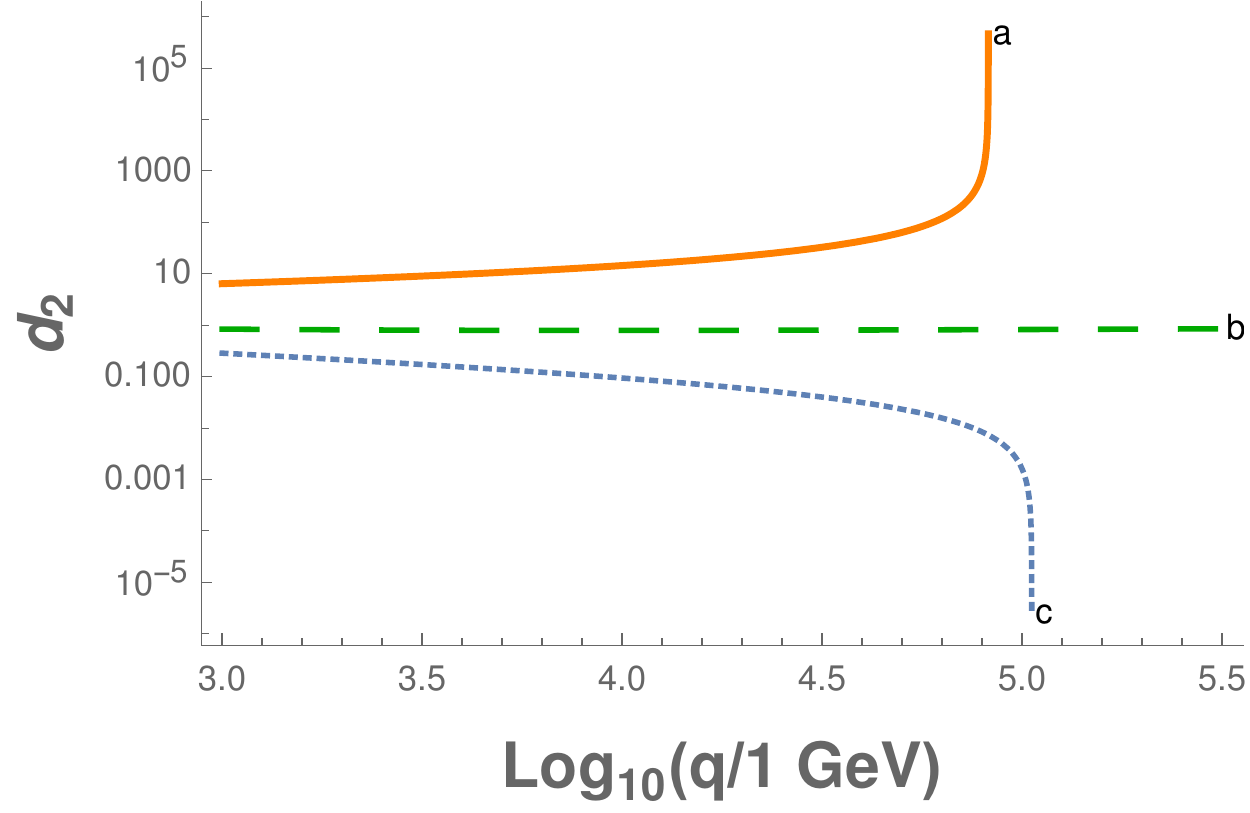} 
\caption{\small{Renormalisation group flow of the scalar quartic coupling $d_2$ in the three models:
CS (left), RSF (centre), and CSF (right). The upper and lower lines in each plot indicate $d_2$ (1 TeV) for which
the model ceases to be valid at 100 TeV, from triviality and stability respectively. The middle dashed line 
indicates a typical value for which the model is stable up to a very high scale, $10^8$--$10^{10}$ GeV. 
}}
\label{fig:rge}
\end{figure}

\begin{table}[htbp]
\begin{center}
\begin{tabular}{ ||c|c||c|c|c|c|c||}
\hline
Model  &   Line   &   $d_2$   &   $\delta_2$ &    $Y$   &   $\alpha$ & $h_Q$ \\
\hline
            &  Upper  &  6.10  &   0.094  &   &  & \\
  CS     & Middle  &  1.21   &  0.042 &  ---  &  0.10 & 0.95 \\
            & Lower  &    0.29   &  0.021 & & & \\
\hline
           &  Upper  &  6.44  &  0.049  &   &   & \\
  RSF   & Middle & 1.00 & 0.019  &  0.85  &  0.05 & 0.68 \\
            &  Lower  & 0.36  &  0.011  &    &   &  \\
 \hline
            &  Upper  &  6.33  &  0.048  & &  &  \\
 CSF   & Middle & 0.83 & 0.017  &  0.70  &  0.05   &   0.95 \\
            &  Lower  & 0.28  &  0.010  &  &    &    \\                    
\hline
\end{tabular}
\caption{Parameter values for the plots of Fig.\ \ref{fig:rge}, with $|q|=2$ for all the 
plots to reproduce the CMS signal strength. The values of $Y$ and $\alpha$ are kept fixed for all the lines 
of any particular model.}
\label{tab:potval}
\end{center}
\end{table}

In Fig.\ \ref{fig:rge}, we show how the scalar quartic $d_2$ is bounded by triviality and stability of the potential
in all the three models. Taking $\Lambda=100$ TeV, up to which we demand the theory to hold, one obtains 
upper and lower limits on $d_2$ at 1 TeV. This, of course, depends on the couplings $\delta_2$ and $Y$ (for
RSF and CSF), as well as the mixing angle $\alpha$\footnote{Triviality and stability bounds on $d_2$ do 
not depend on $\alpha$, but bounds on $\delta_{2}$ do.}, which is shown in Table \ref{tab:potval}. More important is
to note that the limits also depend on the Yukawa coupling $h_Q$, whose value one may extract from the 
CMS signal strength. The values have been chosen in such a way that 
they satisfy $\Omega_{\rm CDM} h^2 \leq 0.12$.

\section{Constraints on the parameter space} \label{sec:results}

The major constraints on the parameter space have been enlisted in Section 2. We assume a thermalised 
dark matter distribution and use micrOMEGAs v5.0.8 \cite{1801.03509}
to obtain the relic density as well as the CDM-nucleon scattering cross-section. The one-loop RG equations
were solved through SARAH v4.14.1 \cite{sarah-staub}, and the unitarity constraints were found from
Ref.\ \cite{goodsell}. The rest of the constraints were dealt with analytically. 

\subsection{The CS model} \label{sec:csanalysis}


\begin{figure}[htbp]
\includegraphics[width=9cm]{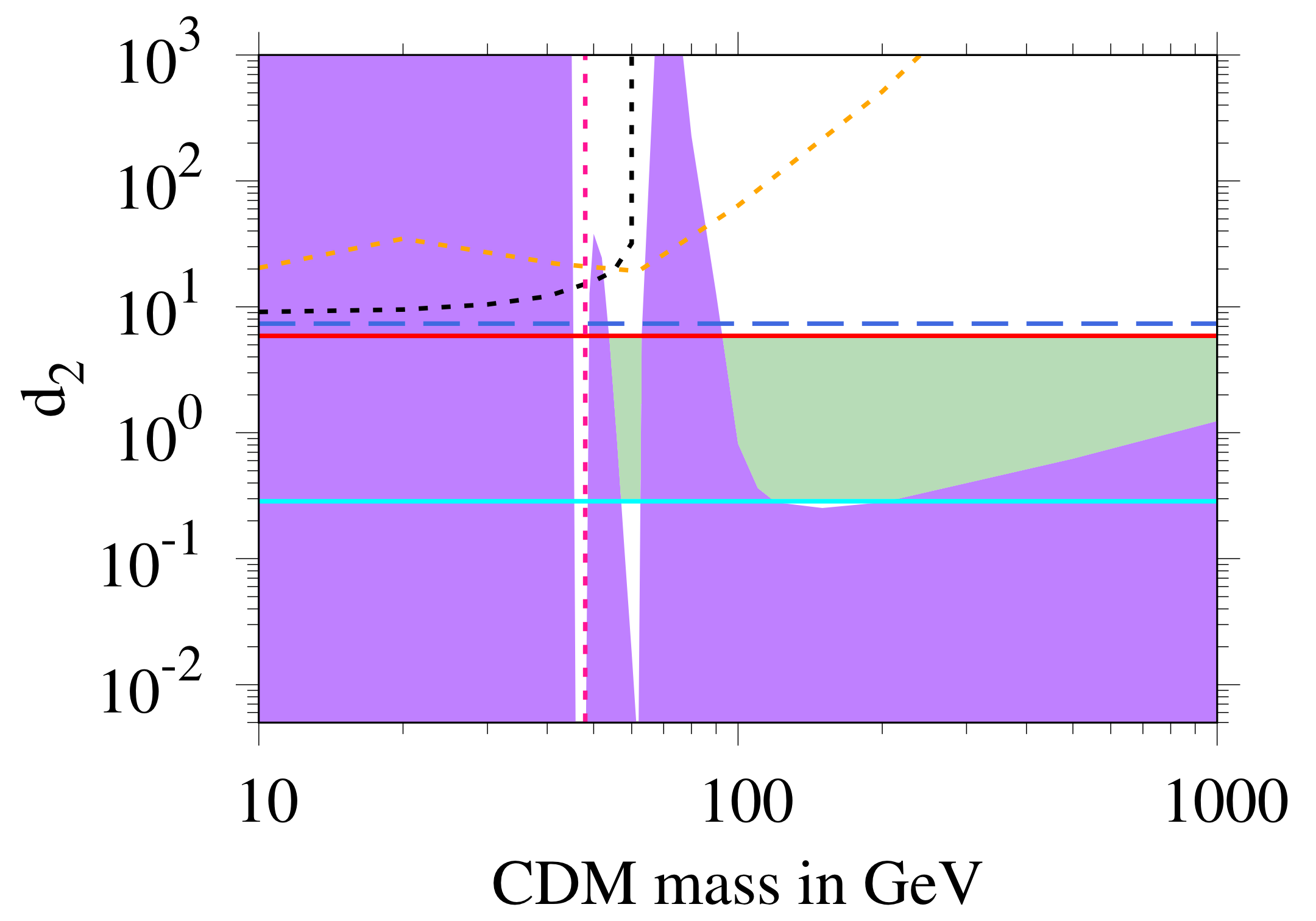} \ \ 
\includegraphics[width=9cm]{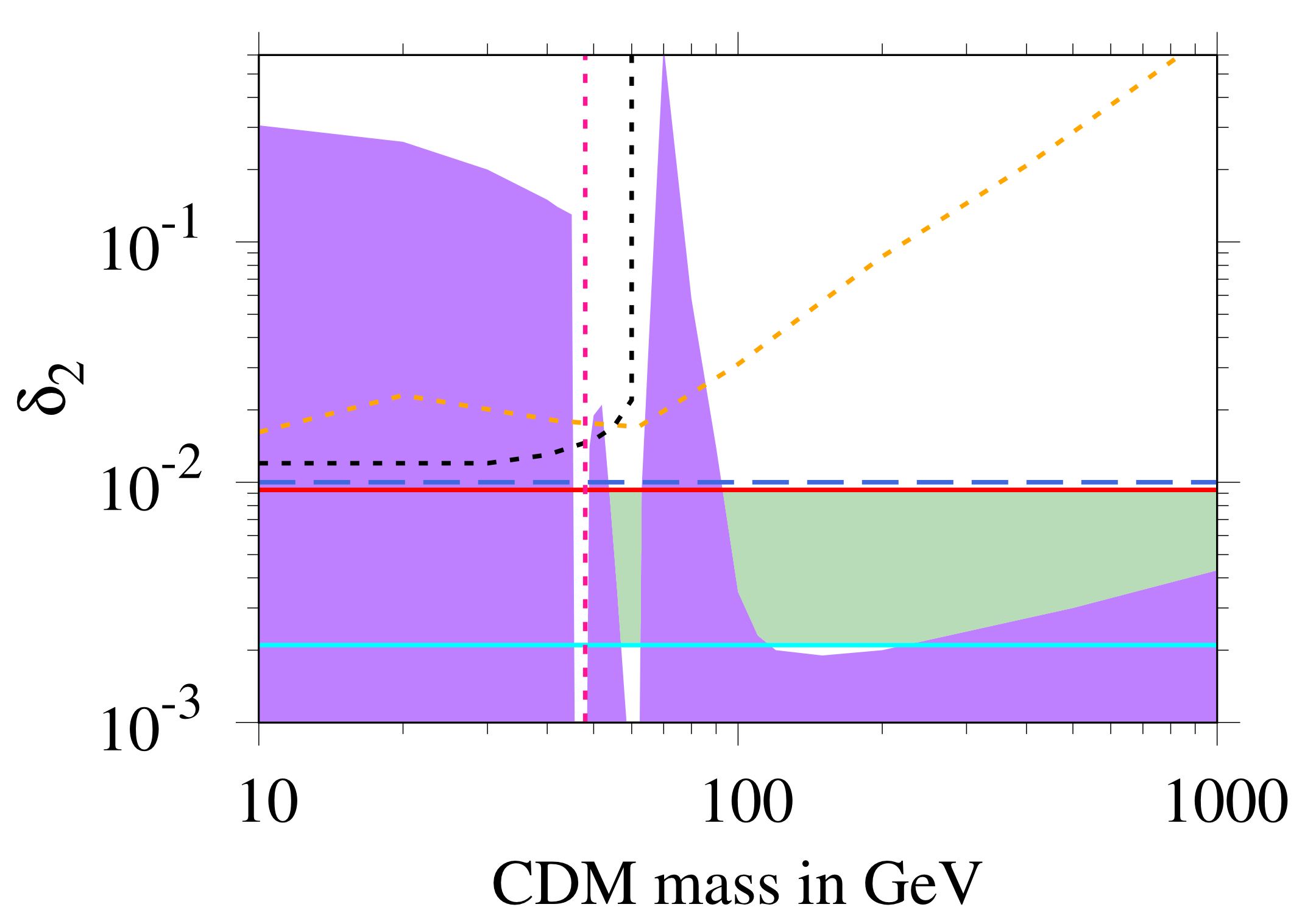}
\caption{\small{The allowed parameter space for the CS model with $\alpha = 0.01$. 
The scalar $S_2$ is the CDM candidate.
The purple region is excluded from overclosure ($\Omega h^2 \geq 0.12$), the region above the dashed black line is 
excluded from the invisible decay of $h$ [BR$(h\to{\rm invis}) < 0.19$], the region above the dashed orange line is 
excluded from the direct detection experiments, the region above the long-dashed blue line is excluded from the 
scattering unitarity constraints, the region above the solid red line is excluded from the triviality 
constraint (no Landau pole before 100 TeV), and the region below the horizontal cyan line is excluded from the 
stability of the potential. The left region of the vertical dashed line at 48 GeV is ruled out 
from the CMS diphoton signal. 
The light green region is allowed from all theoretical and experimental constraints.
}}
\label{fig:cdm-cs1}

\includegraphics[width=9cm]{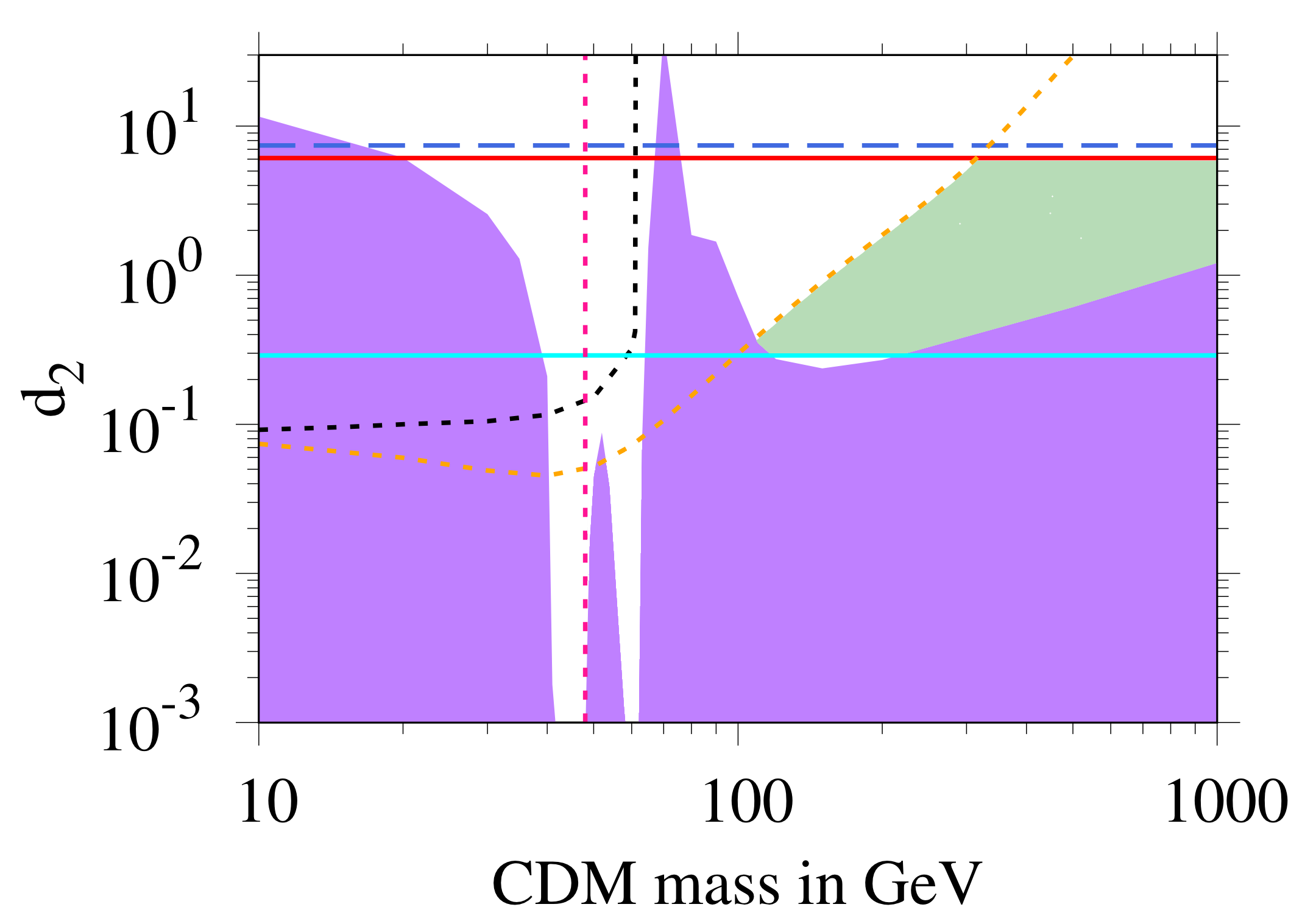} \ \ 
\includegraphics[width=9cm]{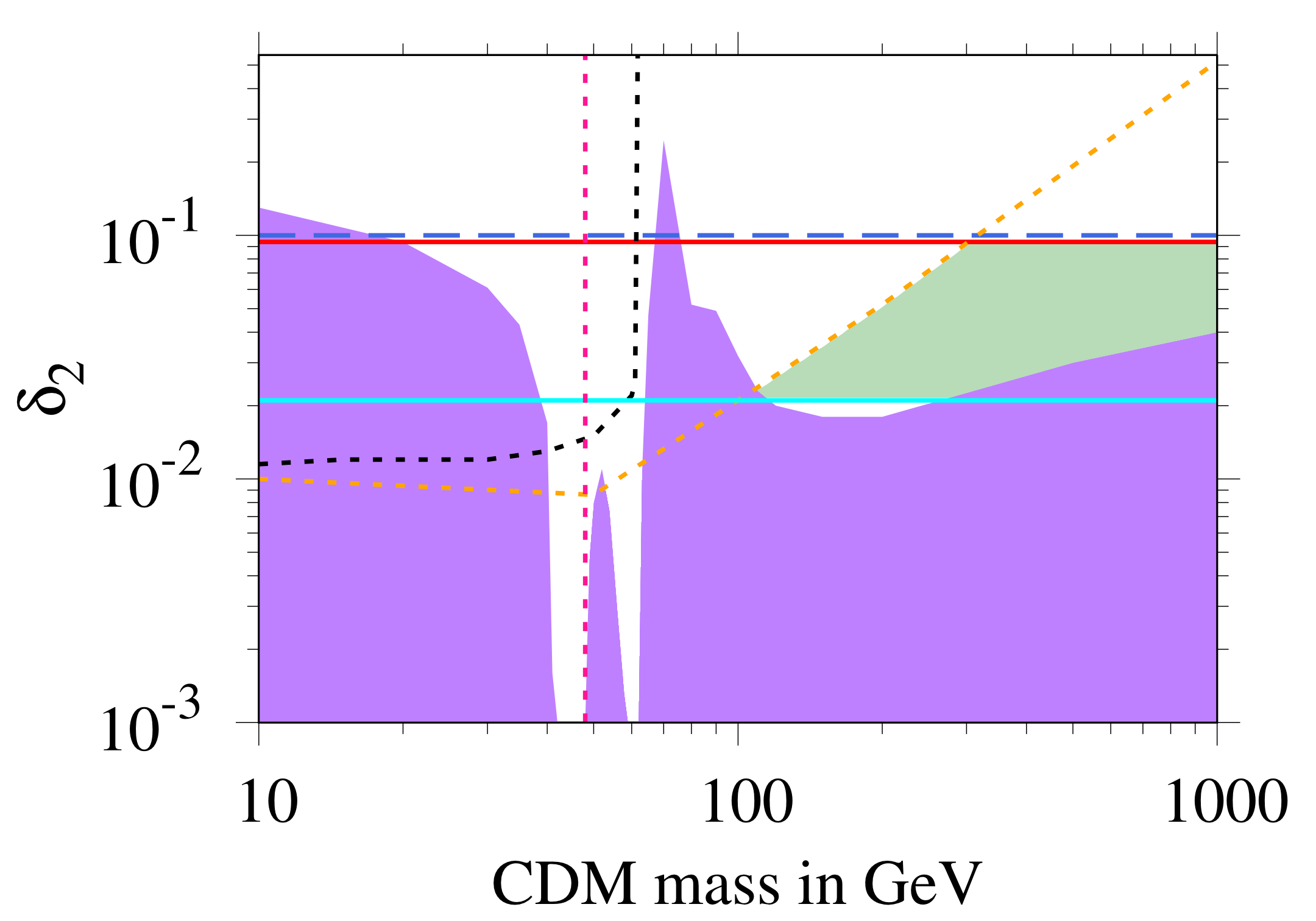}
\caption{\small{The CS parameter space plot as in Fig.\ \ref{fig:cdm-cs1}, but for $\alpha = 0.1$. 
Legends remain the same.}
}
\label{fig:cdm-cs2}
\end{figure}
Let us first explain the strategy for the CS model. As $m_\chi$ is fixed at 96 GeV, only two relevant free
parameters remain: $\alpha$ and $s_1$. From Eq.\ (\ref{eq:reln1}), one can trade them for the scalar quartics 
$d_2$ and $\delta_2$. The third free parameter, $b_1$, is nothing but the mass squared of the CDM, {\em i.e.},
$S_2$. Note that throughout our discussion, we keep $h_Q$ fixed at $0.95$ with $|q|=2$. If the CMS signal 
strength changes, so will $h_Q$, and therefore the horizontal cyan line of all these plots that denotes the 
lower bound on the scalar quartics will shift its position. If the signal disappears altogether, the vertical line at 
$m_{\rm CDM}=48$ GeV will no longer be there; we display the low-mass CDM region keeping that possibility 
in mind.  

The relic density, therefore, is a function of $d_2$, $\delta_2$, and $m_{S_2}$. 
The mixing angle $\alpha$ has to be small enough to maintain the dominantly doublet nature of $h$.
Thus, a good strategy is to fix $\alpha$ and see what values of $s_1$ produce the correct relic density for
different CDM masses $m_{S_2}$. Alternatively, one can show the constraints taking $d_2$ or $\delta_2$ 
as the free parameter.  

\begin{figure}[h]
\includegraphics[width=9cm]{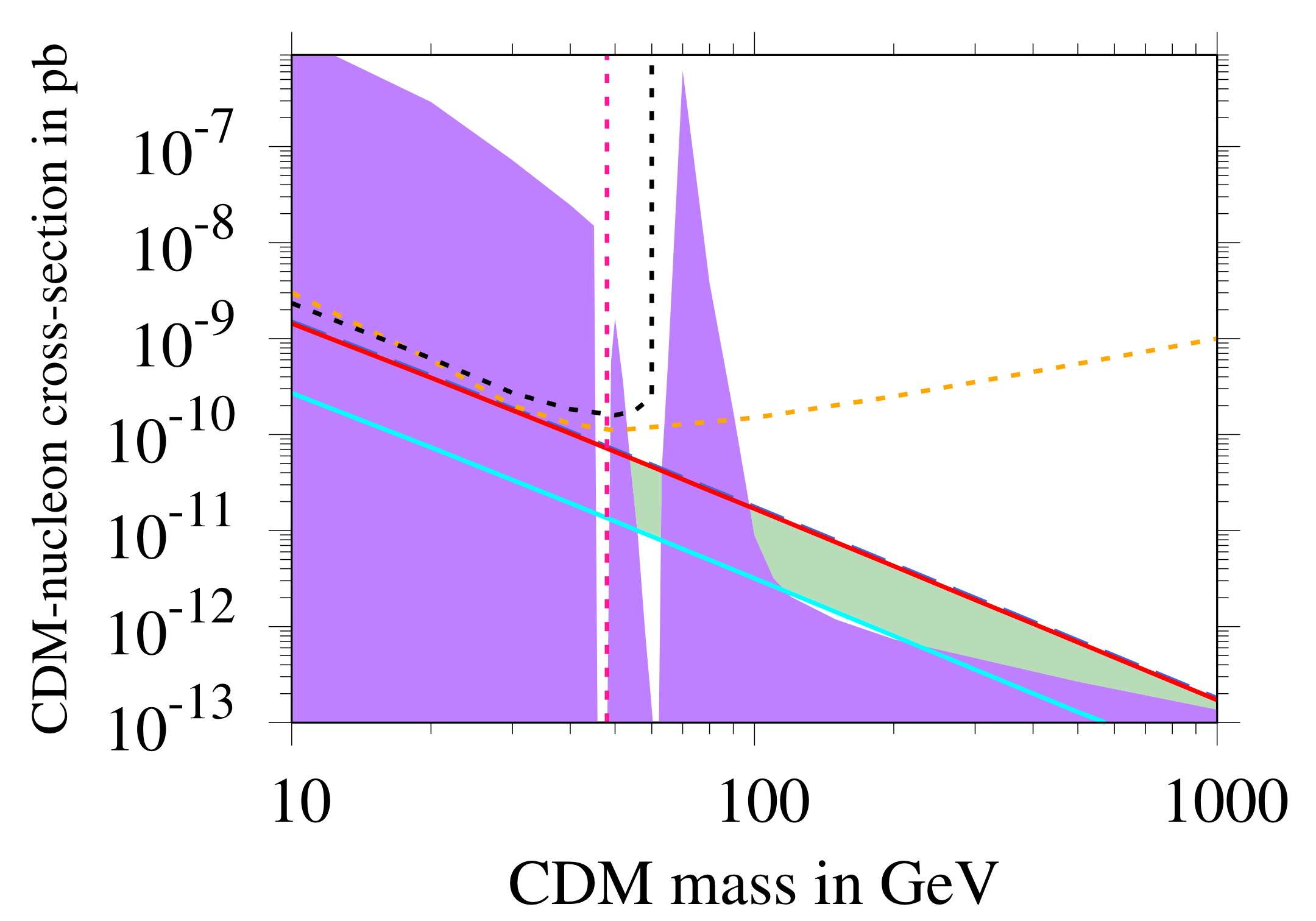} \ \ 
\includegraphics[width=9cm]{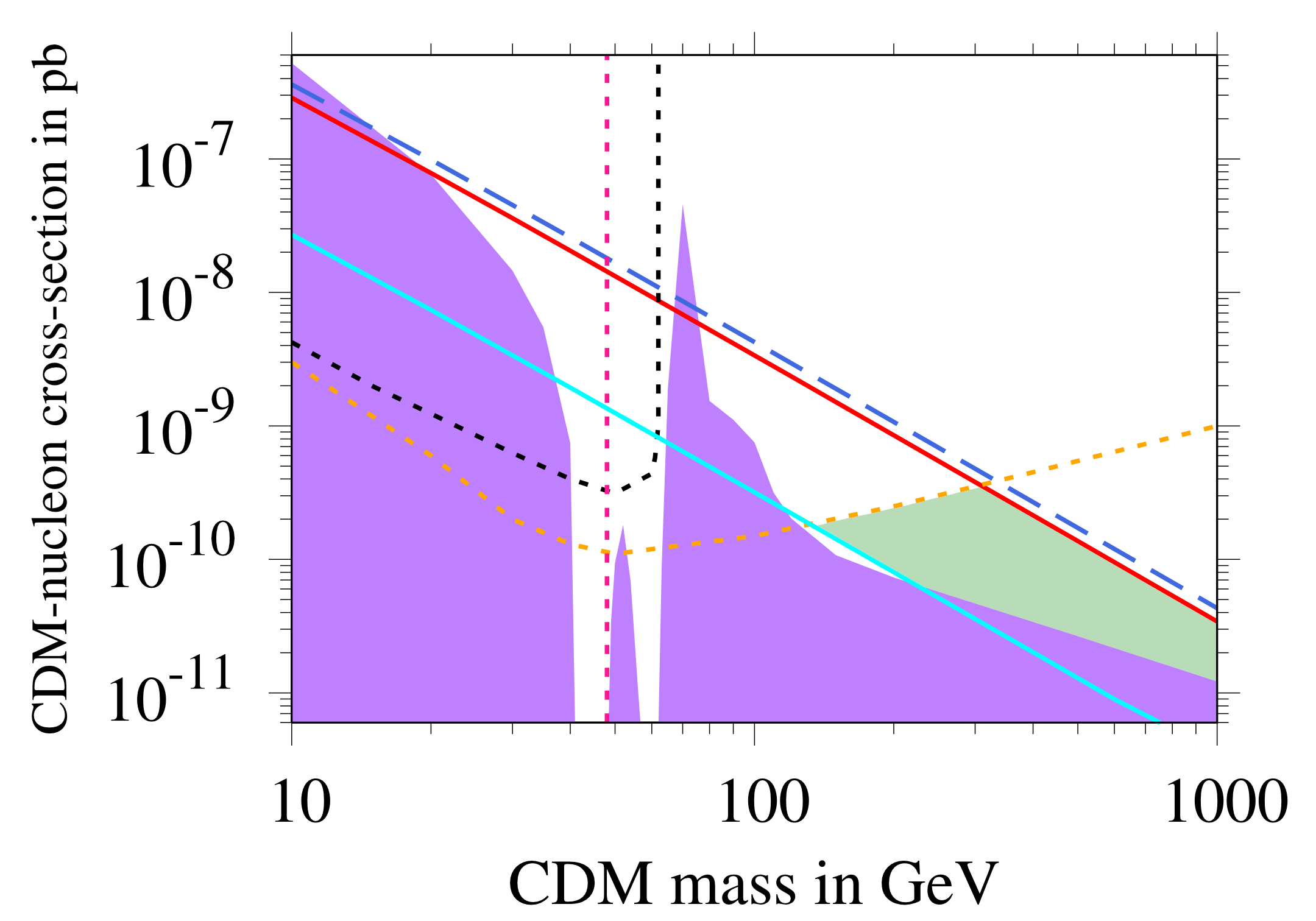}
\caption{\small{The CDM-nucleon cross-section versus the CDM mass for the CS model, for $\alpha = 0.01$ 
(left) and $\alpha = 0.1$ (right). Legends remain the same as in Fig.\ \ref{fig:cdm-cs1}. Only the green shaded 
region is allowed. }
}
\label{fig:cdm-cs3}
\end{figure}

A look at Fig.\ \ref{fig:cdm-cs1} should tell the reader how the constraints work. The left plot displays the 
allowed region for $d_2$, and the right one for $\delta_2$. The green shaded region is what ultimately remains
allowed, after all the theoretical and experimental constraints are imposed. 
If $m_{\rm CDM} < m_\chi/2 \approx 48$ GeV, $\chi$ decays almost entirely to the CDM pair, and each of
the models fail to explain the CMS signal strength, so that region is ruled out, shown by a vertical dashed line
in the subsequent plots. For $m_{\rm CDM} > m_\chi$, the overclosure bound is almost entirely controlled 
by the CDM pair annihilation to $\chi\chi$. Below this, the CDM pair can annihilate to $b\bar{b}$ (through the 
doublet component of $\chi$), $\gamma\gamma$, or $gg$ (through the effective operators). All these 
channels have been taken into account for our analysis.

The salient features of these plots are:
\begin{itemize}

\item The mixed quartic $\delta_2$ is much more tightly constrained from the pure singlet quartic $d_2$.
This is because of its role in the scalar mass matrix as well as the RG equations. However, the nature of 
the two plots is very similar. 

\item For low values of $\alpha$, the parameter space ruled out by the direct detection limit and that ruled out from 
the invisible Higgs decay are competitive. 
However, triviality and scattering unitarity (which essentially put some upper bounds on the couplings) 
rule out a significant amount of parameter space that is still allowed by direct detection; only a narrow 
slice is ruled out by the stability condition. 
Altogether, these constraints rule out a large chunk of the parameter space, particularly for $m_{S_2} > 200$
GeV. Again, this depends on the scale $\Lambda$ where the CS model is taken over by some 
ultraviolet-complete theory. 
The plots were drawn with $\Lambda =100$ TeV; if it is higher, the green region will be even more squeezed. 

\item The single narrow resonance region for the RS model now expands to two closely lying such regions.
The lower mass one comes from the apparent stability of $\chi$, and the higher mass one from that of $h$. 

\item At the same time, the CDM mass below 48 GeV is ruled out from the CMS signal; this is shown by 
a vertical dashed line in Fig.\ \ref{fig:cdm-cs1} and all subsequent figures. In this region, $\chi$ decays almost
entirely to a CDM pair, and the branching ratio to diphoton suffers a huge suppression by several orders 
of magnitude, incompatible with the signal strength. 
\end{itemize}

In Fig.\ \ref{fig:cdm-cs2}, we show identical plots but for $\alpha = 0.1$. While the qualitative features remain
similar to Fig.\ \ref{fig:cdm-cs1}, there are two important changes. 
First, for the $\delta_2$-plot, the direct 
detection limit remains more or less the same, but the stability line has moved up, indicating a tighter lower limit. 
The close pair of lines, marking scattering unitarity and triviality bounds, have also moved up, relaxing the 
corresponding bounds. For the $d_2$-plot, these lines do not change, which 
can qualitatively be understood from the RG equations. What changes significantly is the direct detection 
bound, and as a result, it starts to cut in 
the theoretical constraints for lower values of $m_{S_2}$, approximately $m_{S_2} < 200$ GeV. Second, 
the resonance regions have become wider, which can be explained easily from the $\alpha$-dependence 
of the couplings between physical scalars. 

One can translate these bounds directly to the CDM-nucleon scattering cross-section limits, which is shown 
in Fig.\ \ref{fig:cdm-cs3}, for both $\alpha=0.01$ and $\alpha=0.1$.

\subsection{The RSF model}   \label{sec:rsfanalysis}


\begin{figure}[htbp]
\includegraphics[width=9cm]{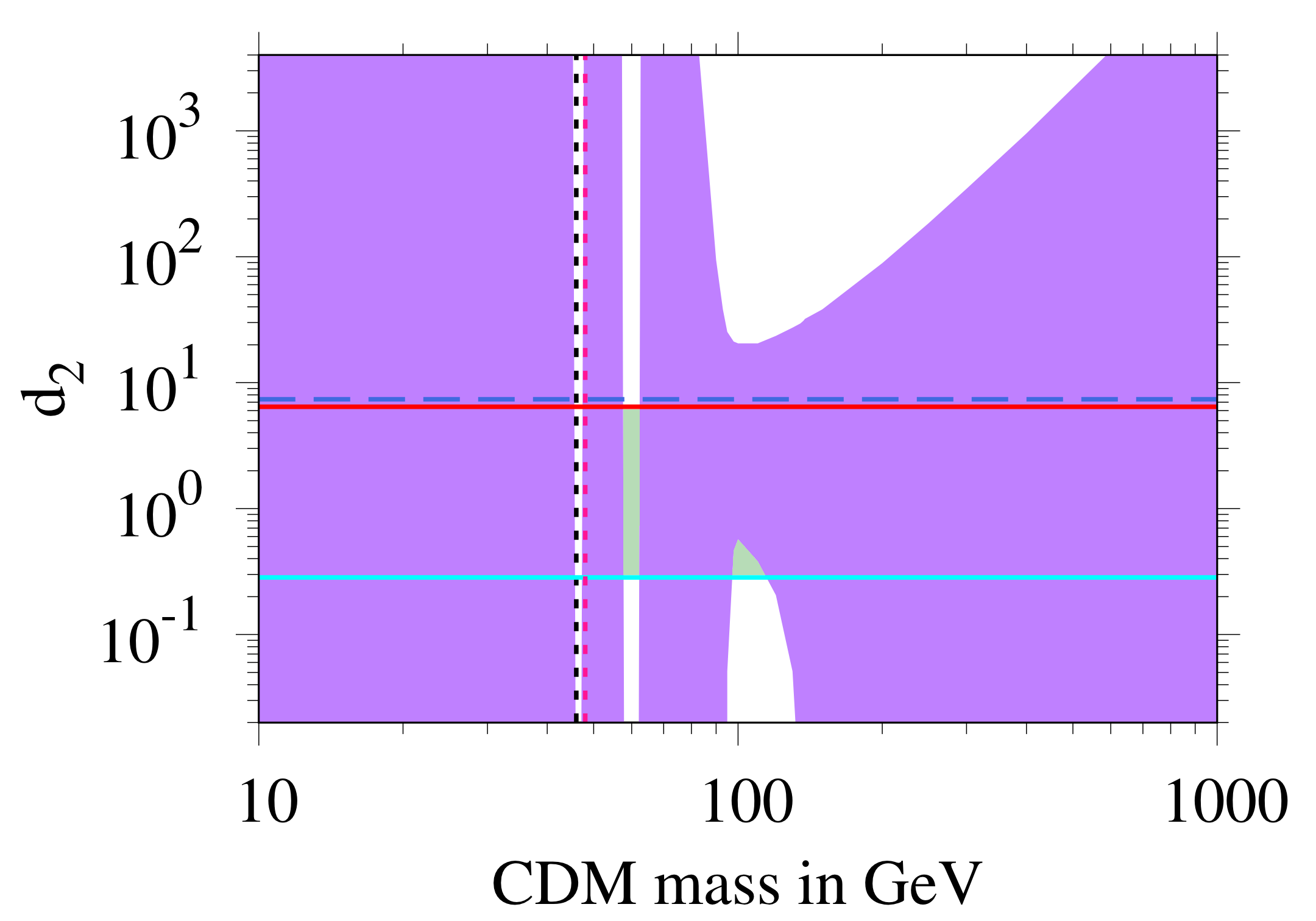} \ \ 
\includegraphics[width=9cm]{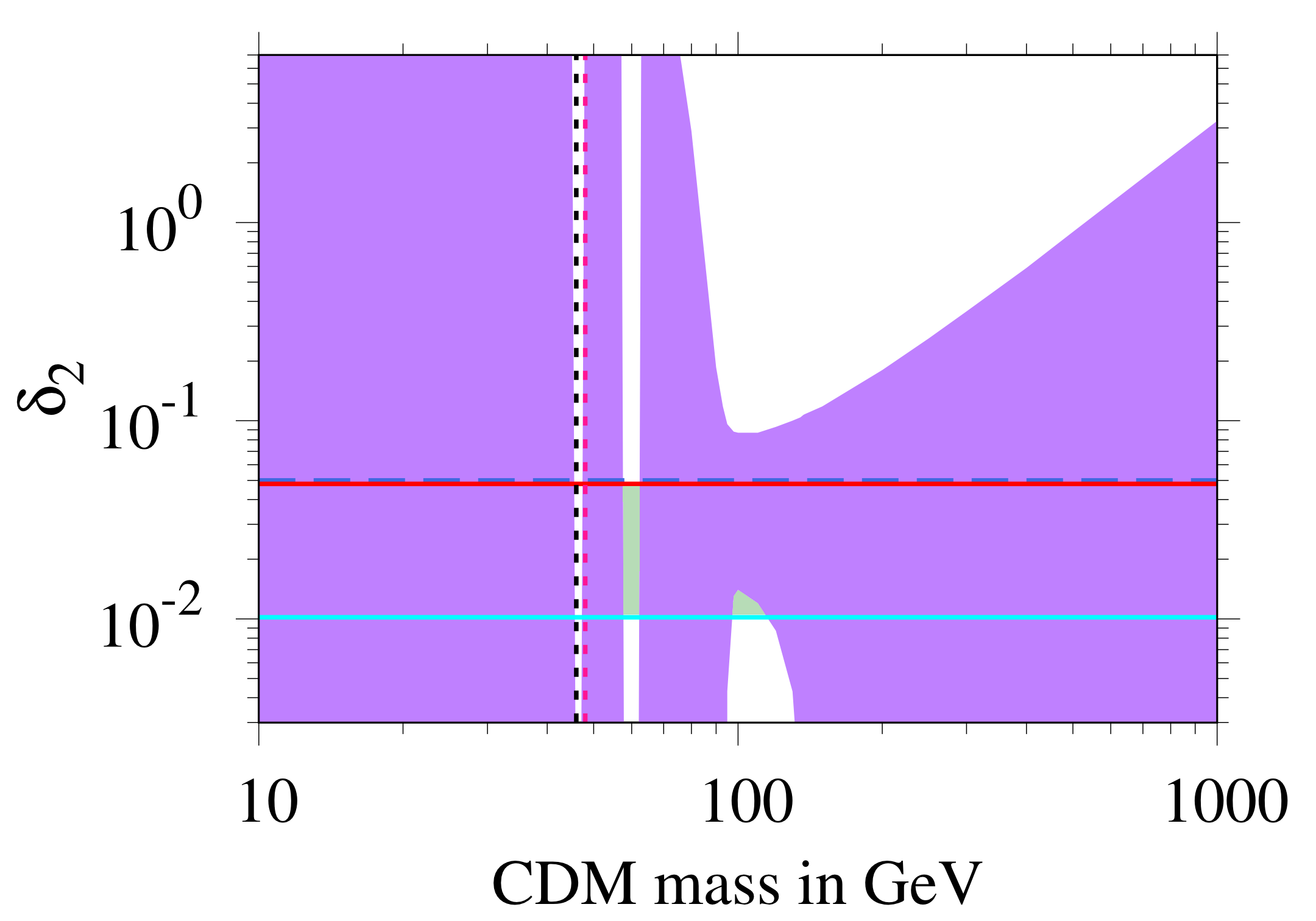}
\caption{\small{The allowed parameter space for the RSF model for $\alpha=0.05$ and Yukawa coupling
$Y=0.7$. The fermion $\psi$ is the CDM candidate.
The purple region is excluded from overclosure, the portion above the red and blue horizontal 
lines (which are overlapping in these plots) are excluded from scattering unitarity and triviality bounds 
respectively, and the portion below the horizontal cyan line near the bottom is excluded from the stability 
of the potential (which becomes unbounded from below). The region left to the short-dashed black line is
excluded from the invisible decay of the Higgs, and that left to the vertical dashed line at 48 GeV from
the CMS signal strength. Only the green shaded region remains allowed. The entire region 
is allowed from the direct detection bounds.}} 
\label{fig:cdm-rsf1}
\includegraphics[width=9cm]{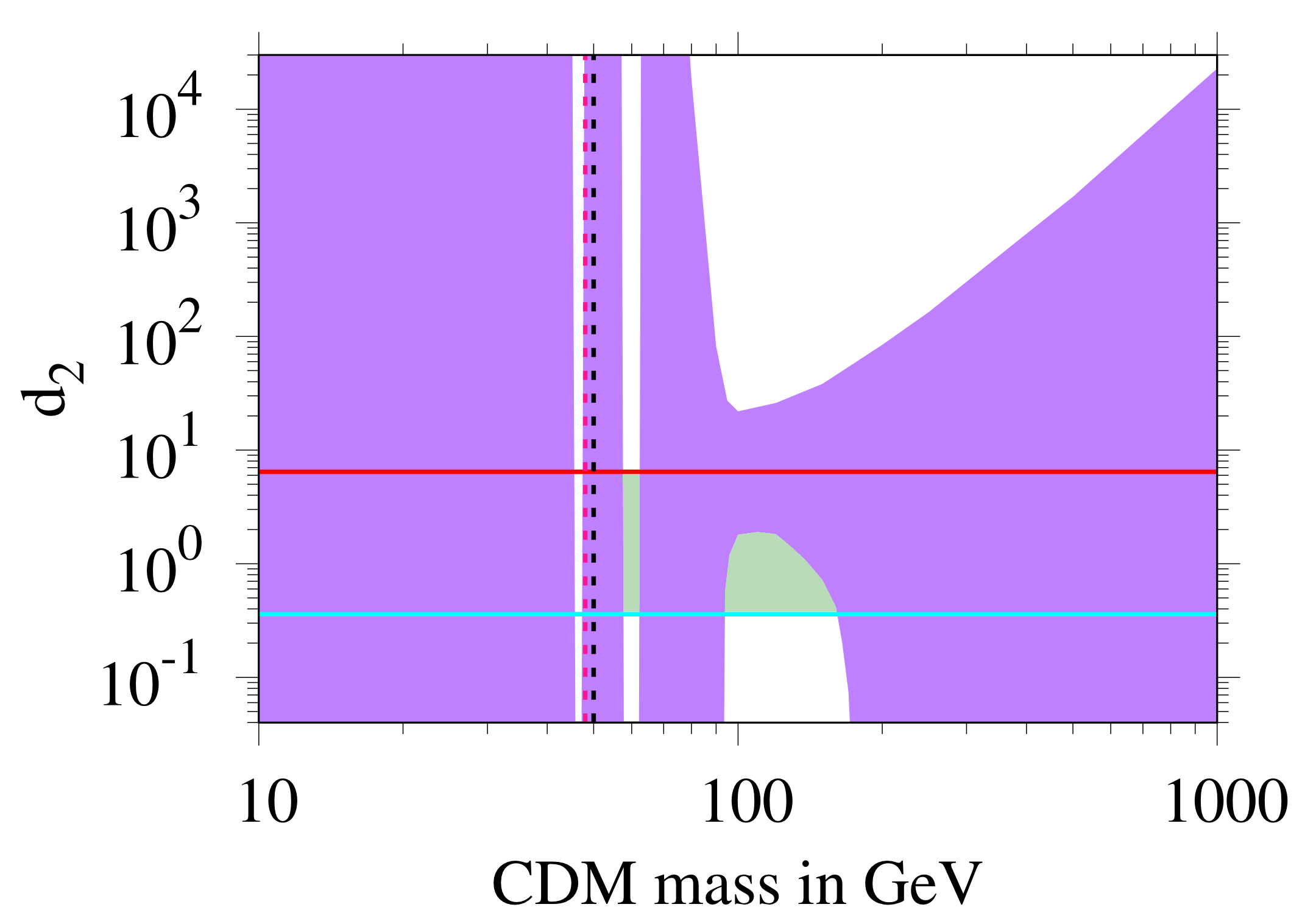} \ \ 
\includegraphics[width=9cm]{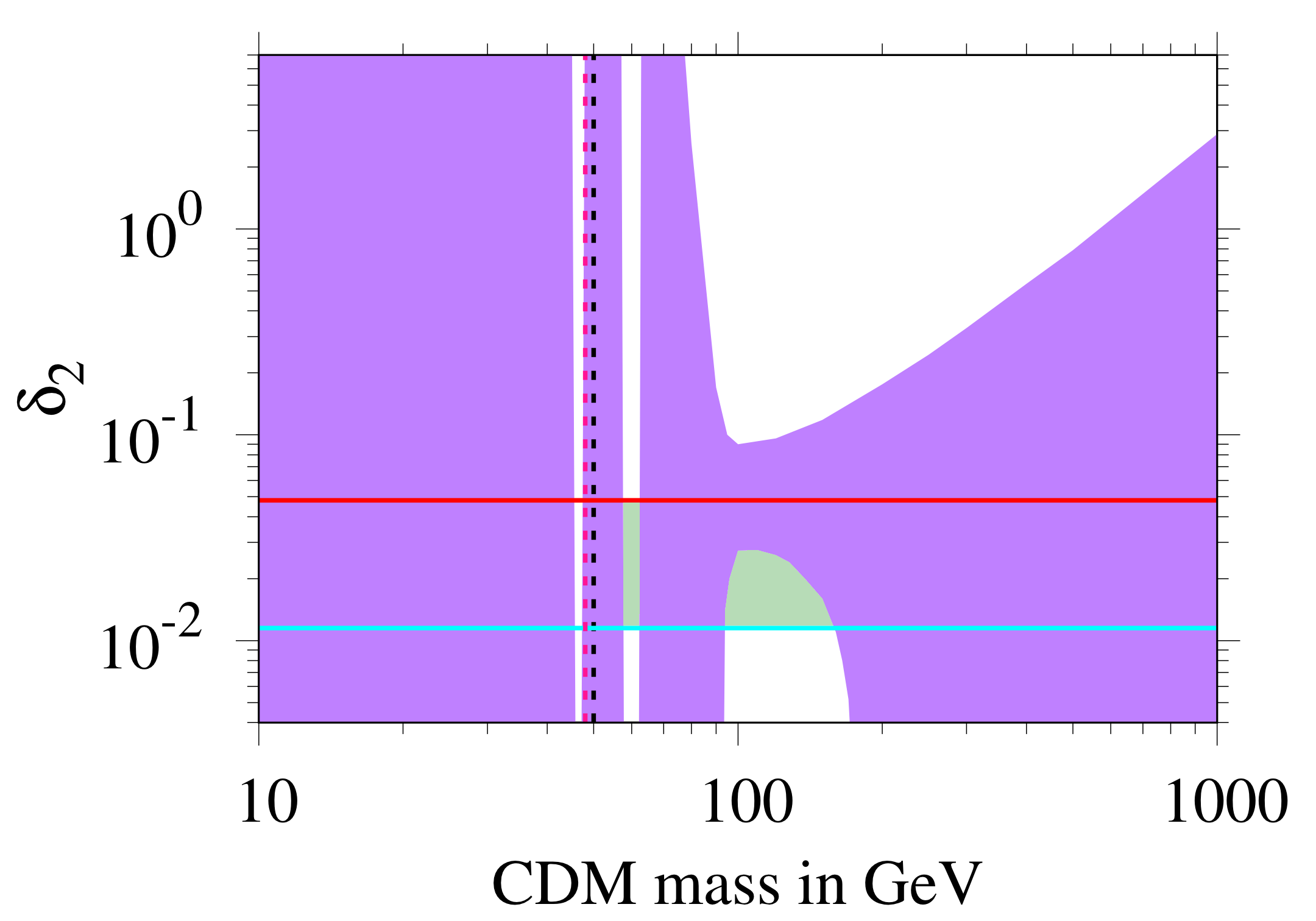}
\caption{\small{The same as Fig.\ \ref{fig:cdm-rsf1} but with $Y=0.85$. All other legends remain the 
same.}}
\label{fig:cdm-rsf2}
\end{figure}

\begin{figure}[h]
\includegraphics[width=9cm]{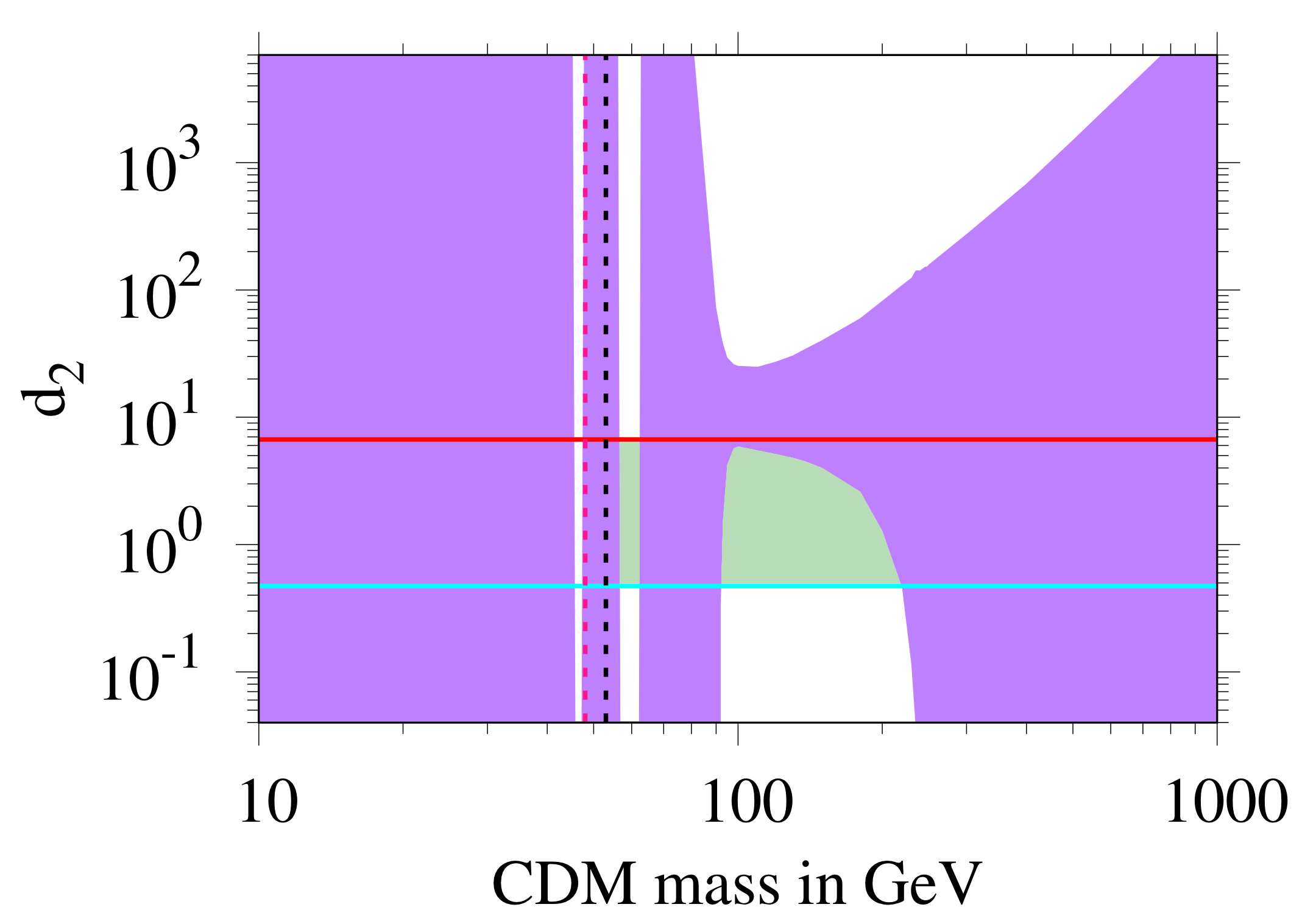} \ \ 
\includegraphics[width=9cm]{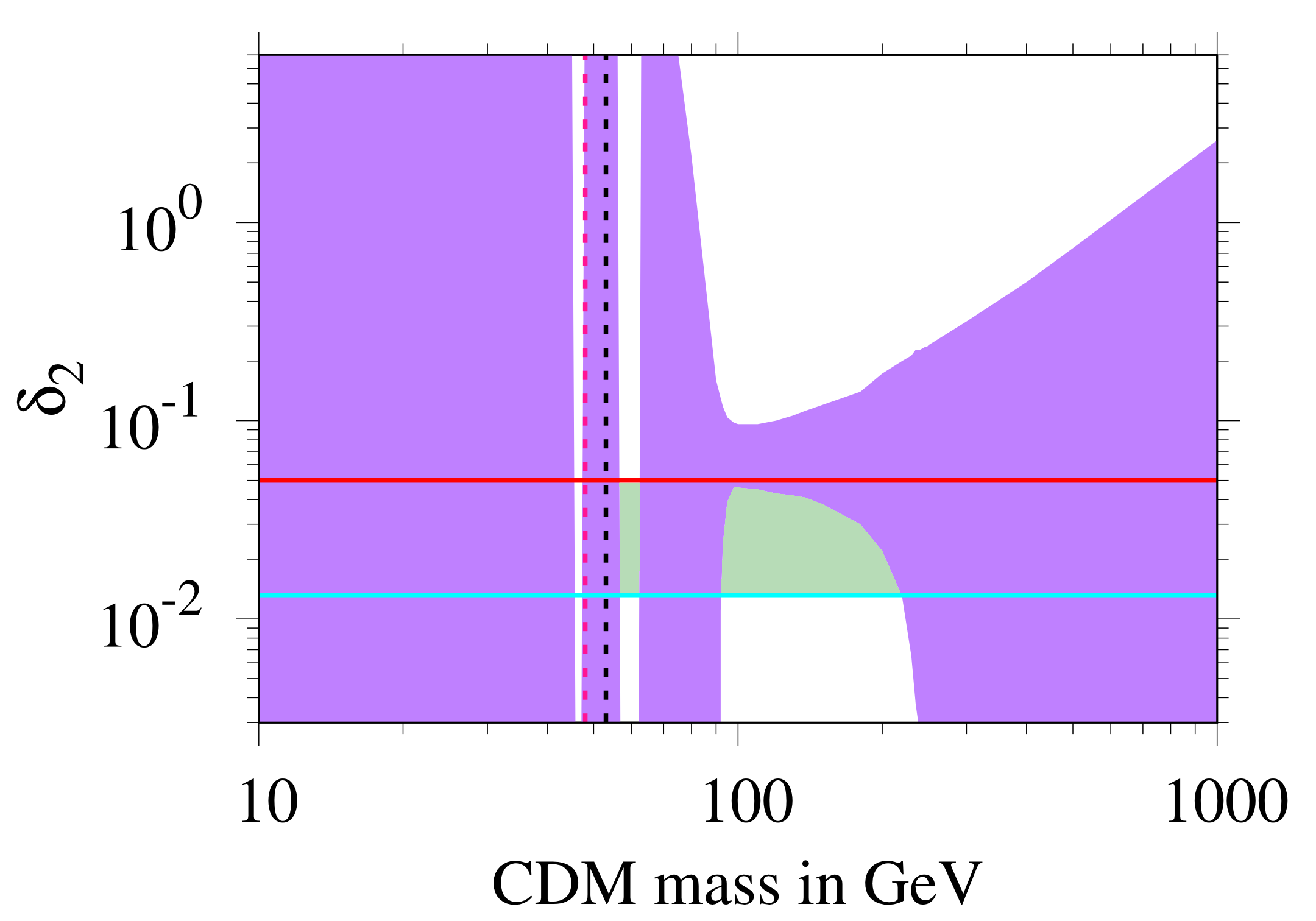}
\caption{\small{The same as Fig.\ \ref{fig:cdm-rsf1} but with $Y=1$. All other legends remain the 
same.}}
\label{fig:cdm-rsf3}
\end{figure}

The allowed parameter space for the RSF model is shown in Figs.\ \ref{fig:cdm-rsf1}, \ref{fig:cdm-rsf2}, and
\ref{fig:cdm-rsf3}. The 
fermion $\psi$ is the CDM candidate, and couples only to the real singlet $S$. Thus, for a large enough 
Yukawa coupling $Y$ between $S$ and $\psi$, or $h_Q$ between $S$ and $Q$ (in Subclass (a) models)
 the potential runs the risk of being unbounded from below at a high
scale. In Figs.\ \ref{fig:cdm-rsf1}-\ref{fig:cdm-rsf3}, we have also delineated the region where such a catastrophe 
takes place (the lower part of the horizontal cyan line). The instability threshold is taken, again, to be at 100 TeV 
or more, {\em i.e.}, some other new physics comes into play at that scale to make the potential stable. While the entire 
region showed in these figures is allowed from the direct detection limits, one may note the power of the 
triviality and stability constraints, which ultimately leaves only a narrow region around 150 GeV as allowed. The 
allowed region expands for larger values of $Y$, as the overclosure bound starts shrinking. 

For Figs.\ \ref{fig:cdm-rsf1}-\ref{fig:cdm-rsf3}, the mixing angle $\alpha$ is fixed at $0.05$. 
For higher values of $\alpha$, $\psi$ couples with $h$ more strongly and hence direct detection constraints 
tend to rule out more of the allowed parameter space
shown here. For even smaller values of $\alpha$, resonance regions no longer exist. It is also easy to 
understand why the constraint from the invisible decay of the Higgs is a vertical line. The invisible decay width 
depends on the CDM mass, $\alpha$, and $Y$, but here $\alpha$ and $Y$ are fixed, 
so the only dependence is on $m_{\rm CDM}$.   

The window near $m_{\rm CDM} \sim 150$ GeV, in Figs.\ \ref{fig:cdm-rsf1}-\ref{fig:cdm-rsf3}, 
is interesting. Let us see what happens when we keep $m_{\rm CDM}$ 
fixed in that region but increase $d_2$ or $\delta_2$. Such a shift decreases $s_1$, the VEV of $S$. This 
increases the relic density to such a point as to hit the overclosure bound, so the region becomes 
disallowed and the neck begins. If we continue increasing $d_2$ or $\delta_2$, the relic density starts 
decreasing after a point ({\em e.g.}, one may look at Fig.\ \ref{fig:cdm-rsf0} to see how the annihilation 
cross-section changes, moving from right to left), and the parameter space again opens up, terminating the neck. 

\begin{figure}[h]
\centerline{ \includegraphics[angle=0,origin=c,width=9cm]{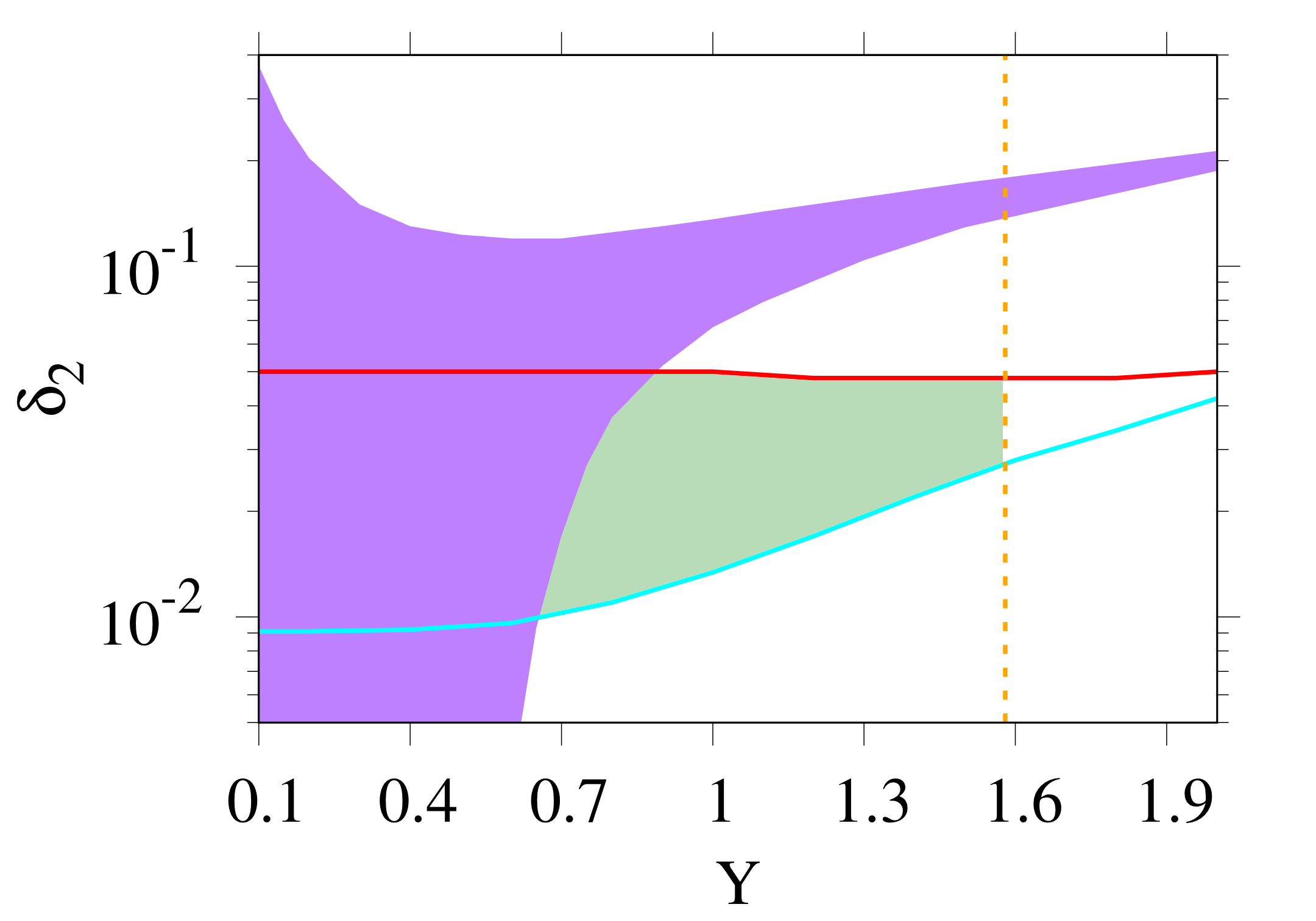} \ \ 
 \includegraphics[angle=0,origin=c,width=9cm]{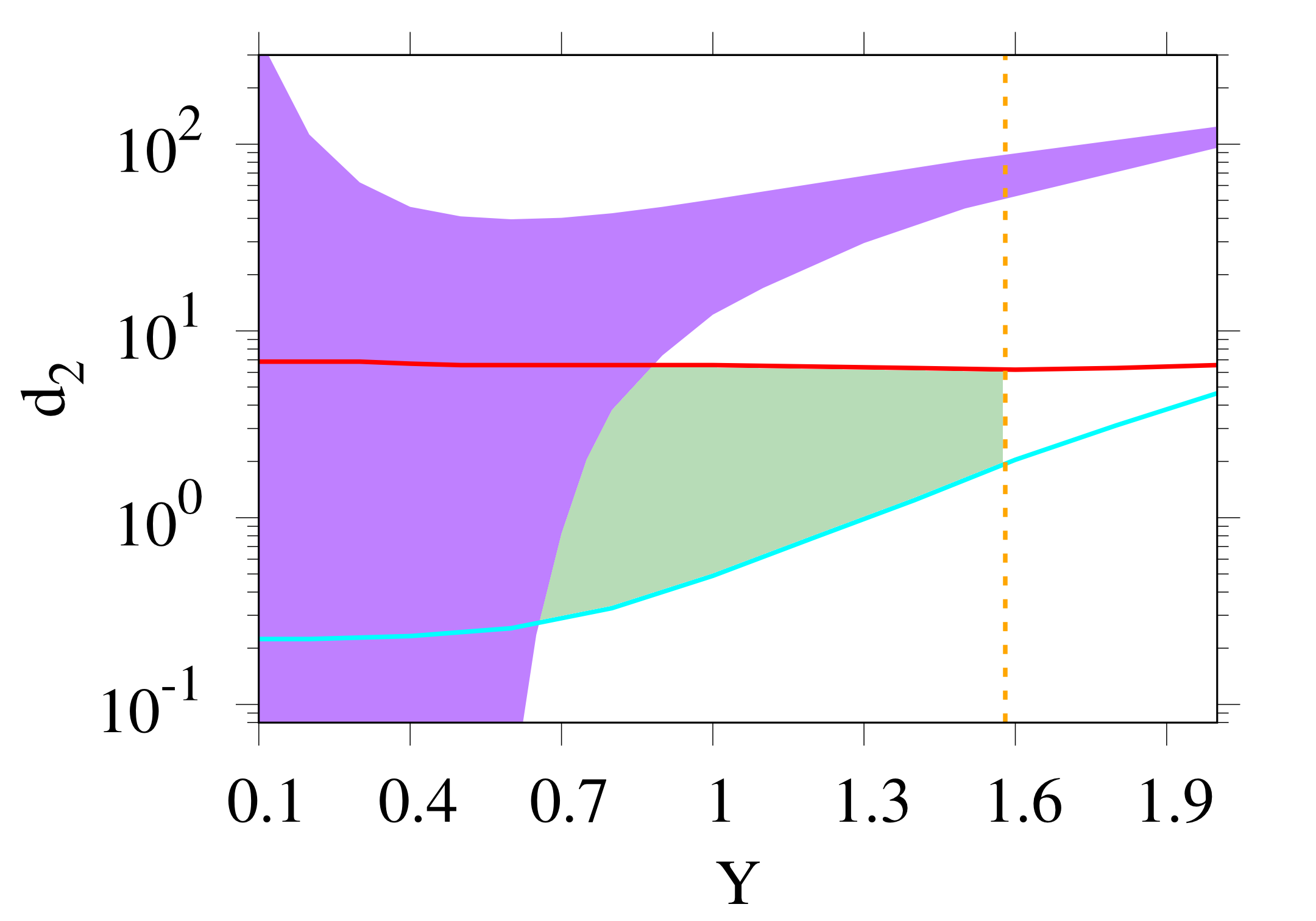}
 }

\caption{\small{
The allowed region for the RSF model, with $\alpha = 0.05$ and $m_{\rm CDM} = 100$ GeV. 
For details, see text. The region below the red horizontal line, above the cyan line, 
right to the overclosure patch, and left to the dashed orange vertical line, is allowed.
}}
\label{fig:cdm-rsf4}
\end{figure}

Let us also display the interdependence of the couplings $d_2$, $\delta_2$, and $Y$ in Fig.\ \ref{fig:cdm-rsf4}. 
Technically, this should depend on $h_Q$ too, but we have used a fixed value of the coupling. 
We kept the mixing angle at $\alpha = 0.05$ and changed the scalar VEV $s_1$, which in turn controls 
$d_2$ and $\delta_2$. Apart from the shaded overclosure region, there are three lines in each plot. The vertical line 
at the right is the bound from direct detection; as the CDM mass is kept fixed at 100 GeV, the limit depends only 
on $Y$, and not on the scalar quartics. The region above the almost horizontal red line is ruled out 
because the couplings blow up before 100 TeV. Large values of $Y$ can delay the onset of the Landau pole; 
that is why the line veers slightly upwards for large $Y$ (although that region is ruled out from direct detection). 
The region below the cyan line is ruled out because the potential, at least at the tree-level, becomes unstable there. 
Only the island within these three lines and the overclosure region remains allowed.

\subsection{The CSF-2 model}
The CSF-1 model, with the fermion being the sole dark matter candidate, is qualitatively very similar to the 
RSF model, even more so if the second singlet scalar is heavy. We, thus, do not explicitly show the features, and 
rather concentrate on CSF-2. 
As both the scalar $S_2$ and the fermion $\psi$ can be CDM candidates, we denote their masses by
$m^S_{\rm CDM}$ and $m^F_{\rm CDM}$ respectively.

\begin{figure}[htbp]
\centerline{ \includegraphics[origin=c,width=9cm]{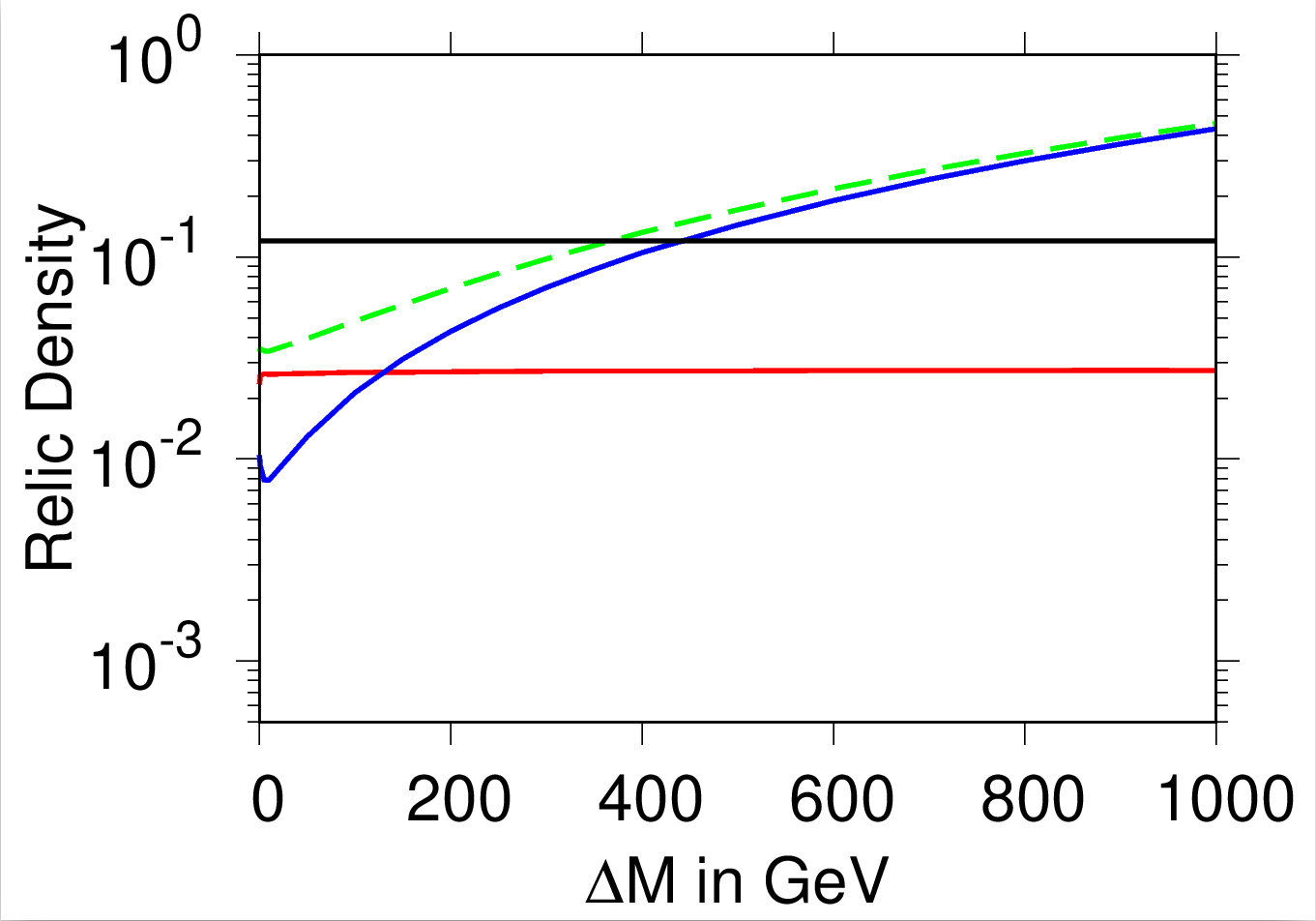} \ \ \ 
 \includegraphics[origin=c,width=9cm]{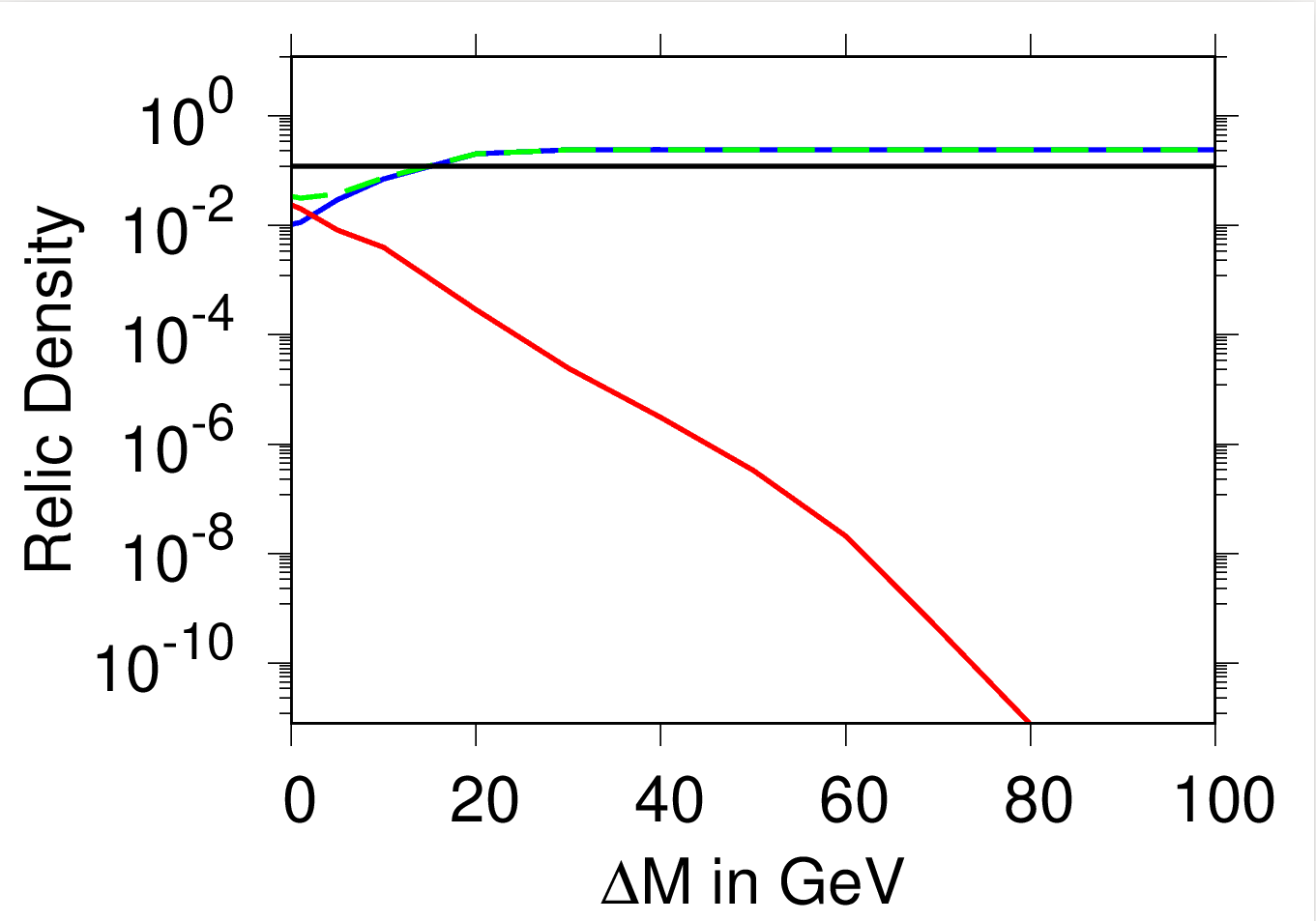}
}
\caption{\small{The composition of the CDM in the CSF-2 model, with the mass difference 
$\Delta M$ (in GeV) of the CDM components plotted against the relic density. The blue line corresponds 
to the contribution of $\psi$ and the red line to that of $S_2$, while the green line gives the combined 
contribution. Left: $m^F_{\rm CDM} > m^S_{\rm CDM} = 100$ GeV.
Right: $m^S_{\rm CDM} > m^F_{\rm CDM} = 100$ GeV.
For both the plots, the Yukawa coupling $Y=0.7$ and the singlet scalar VEV $s_1=100$ GeV. 
The relic density limit $\Omega h^2 = 0.12$ is also shown by the black horizontal line.
}}
\label{fig:cdm-csf1}
\includegraphics[width=9cm]{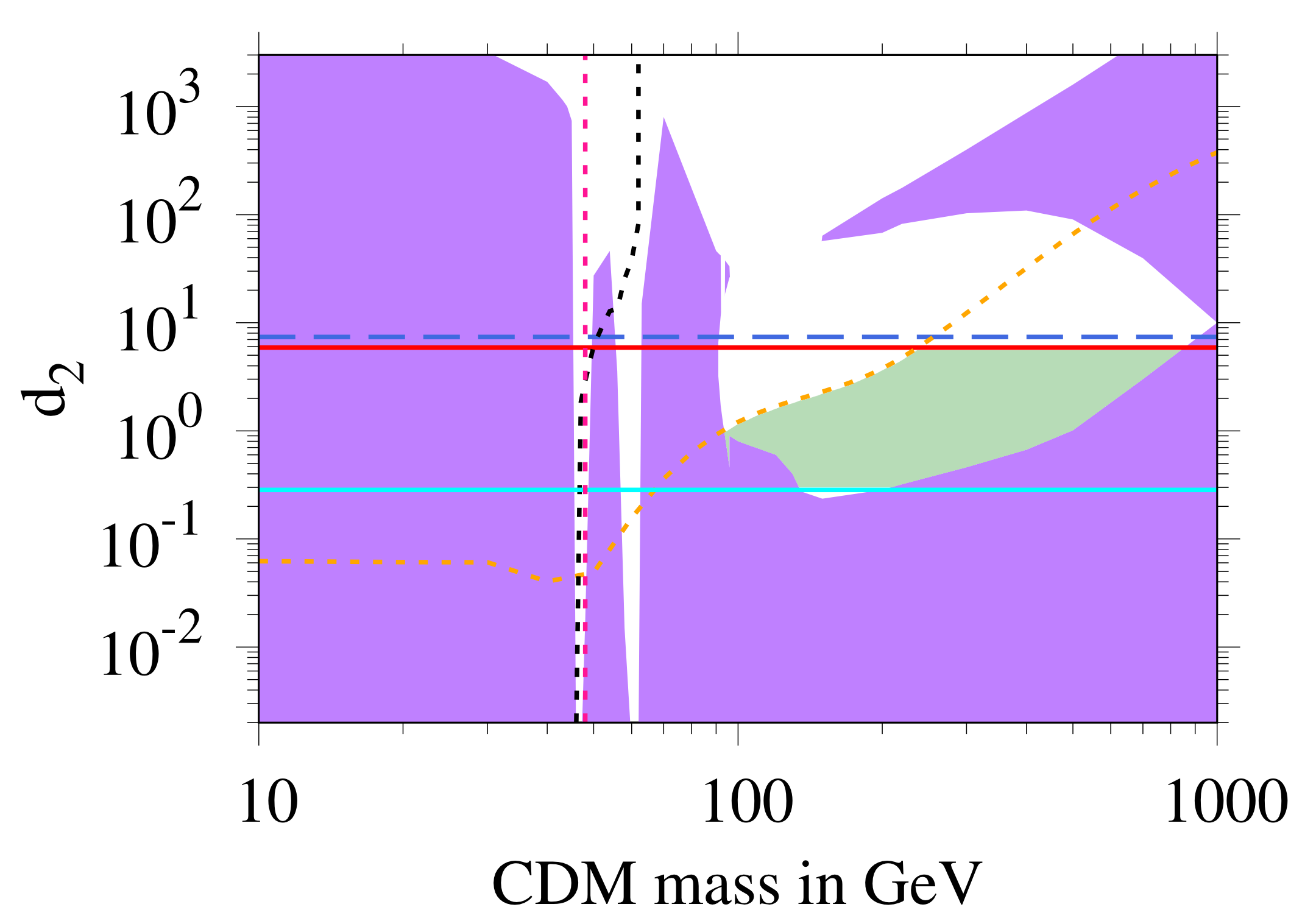} \ \ 
\includegraphics[width=9cm]{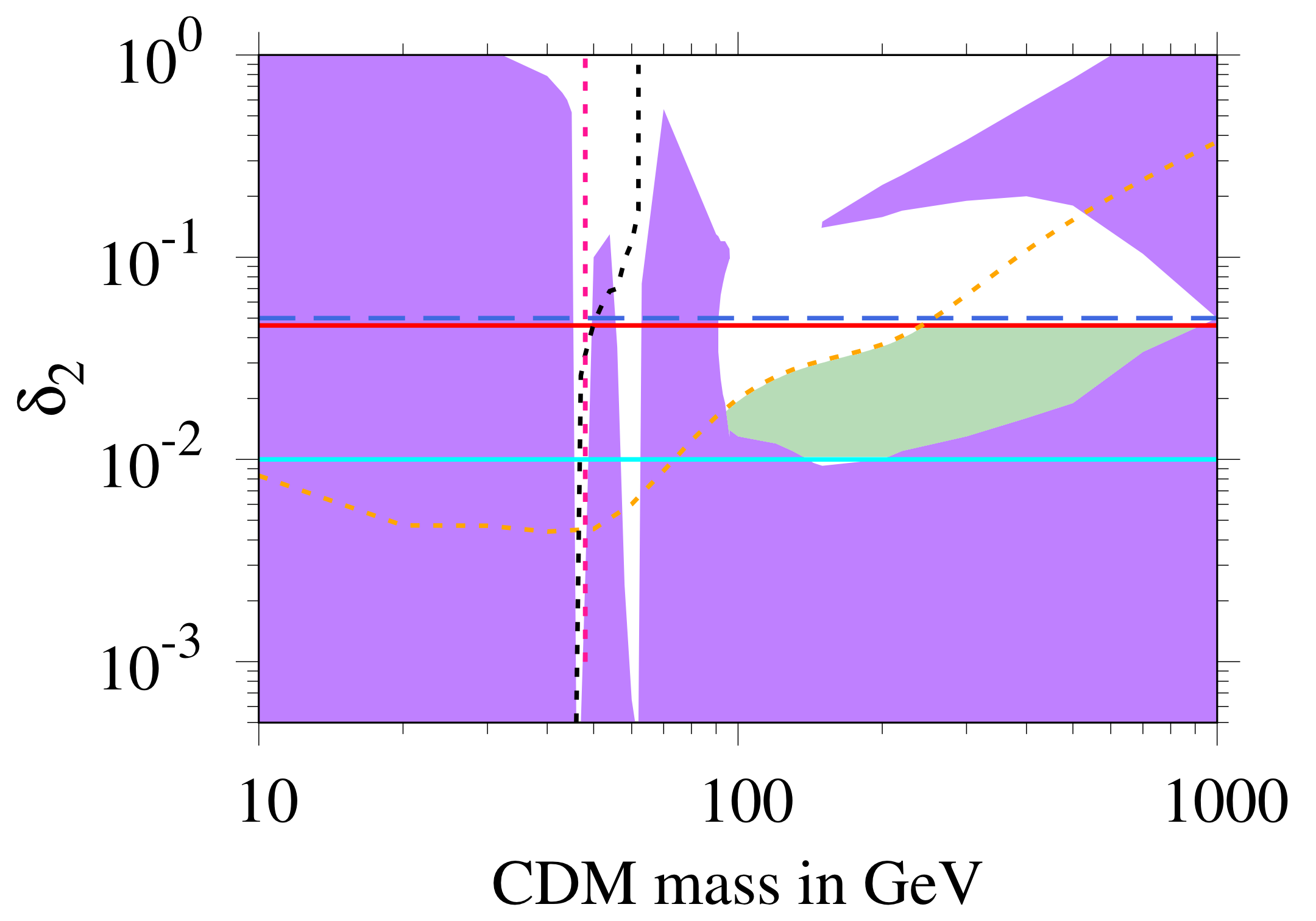}
\caption{\small{The allowed parameter space for the CSF-2 model for $\alpha=0.05$ and Yukawa coupling
$Y=0.7$. The legends are identical with that of Fig.\ \ref{fig:cdm-rsf1}, apart from the dashed orange 
line, the region above which is excluded from the direct detection experiments. For more explanation,
see text.
}}
\label{fig:cdm-csf2}
\end{figure}

\begin{figure}[h]
\centerline{\includegraphics[width=9cm]{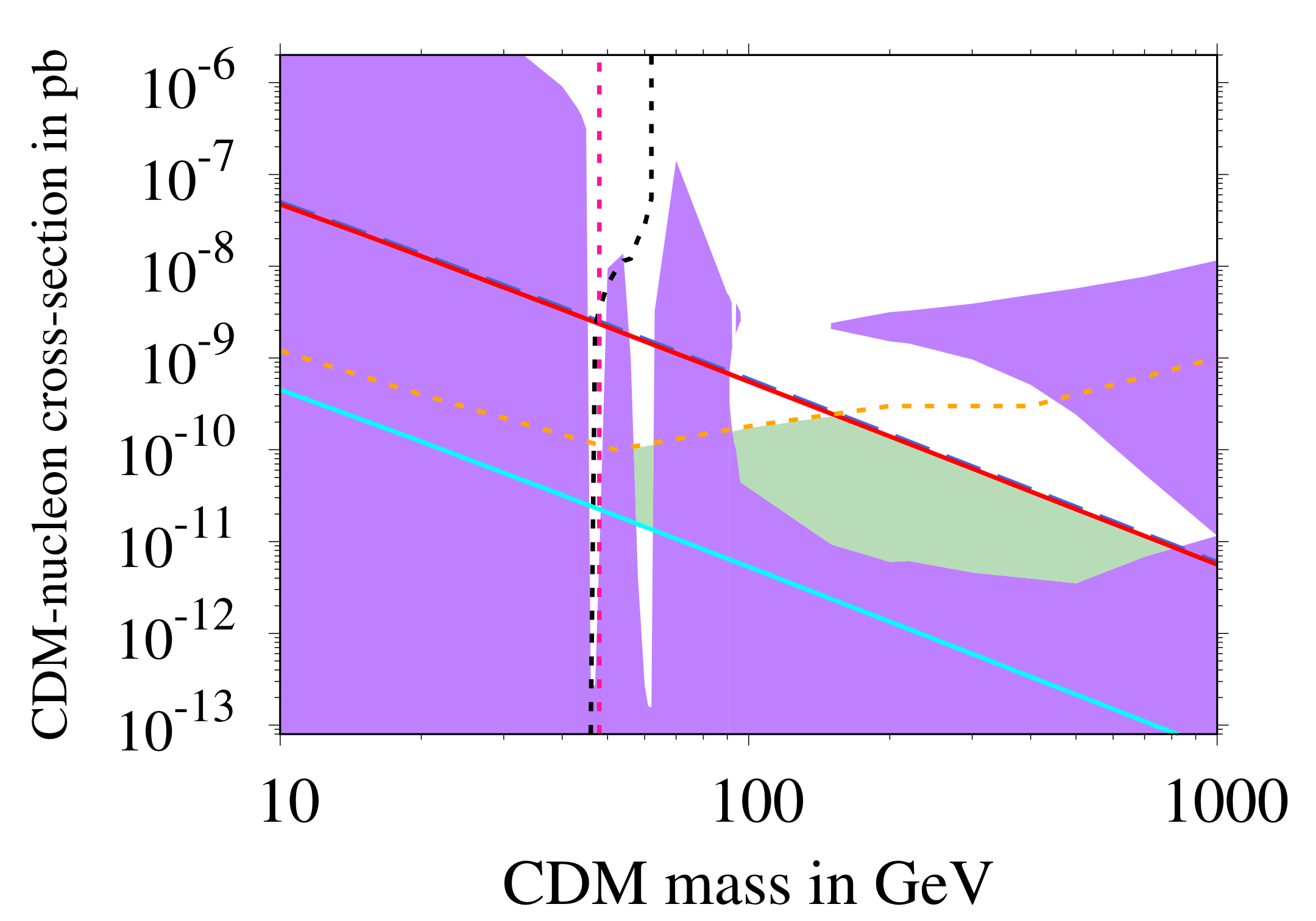} }
\caption{\small{The CDM-nucleon cross-section versus the CDM mass for the CSF-2 model for $\alpha = 0.05$ and Yukawa coupling $Y=0.7$. Legends remain the same as in Fig.\ \ref{fig:cdm-csf2}. Only the green shaded 
region is allowed. }
}
\label{fig:cdm-csf3}
\end{figure}

\begin{figure}[h]
\centerline{ \includegraphics[angle=0,origin=c,width=9cm]{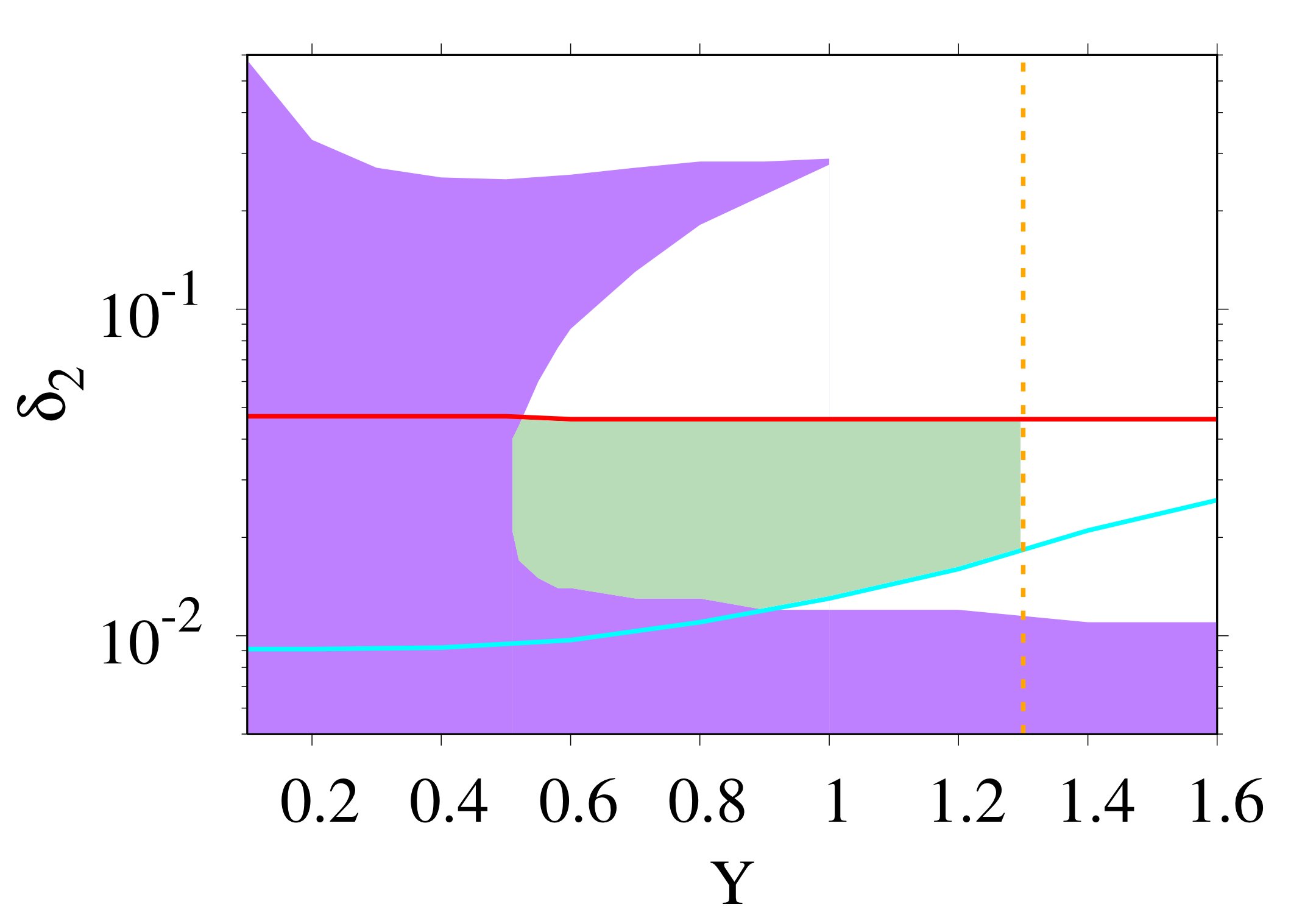} \ \ 
 \includegraphics[angle=0,origin=c,width=9cm]{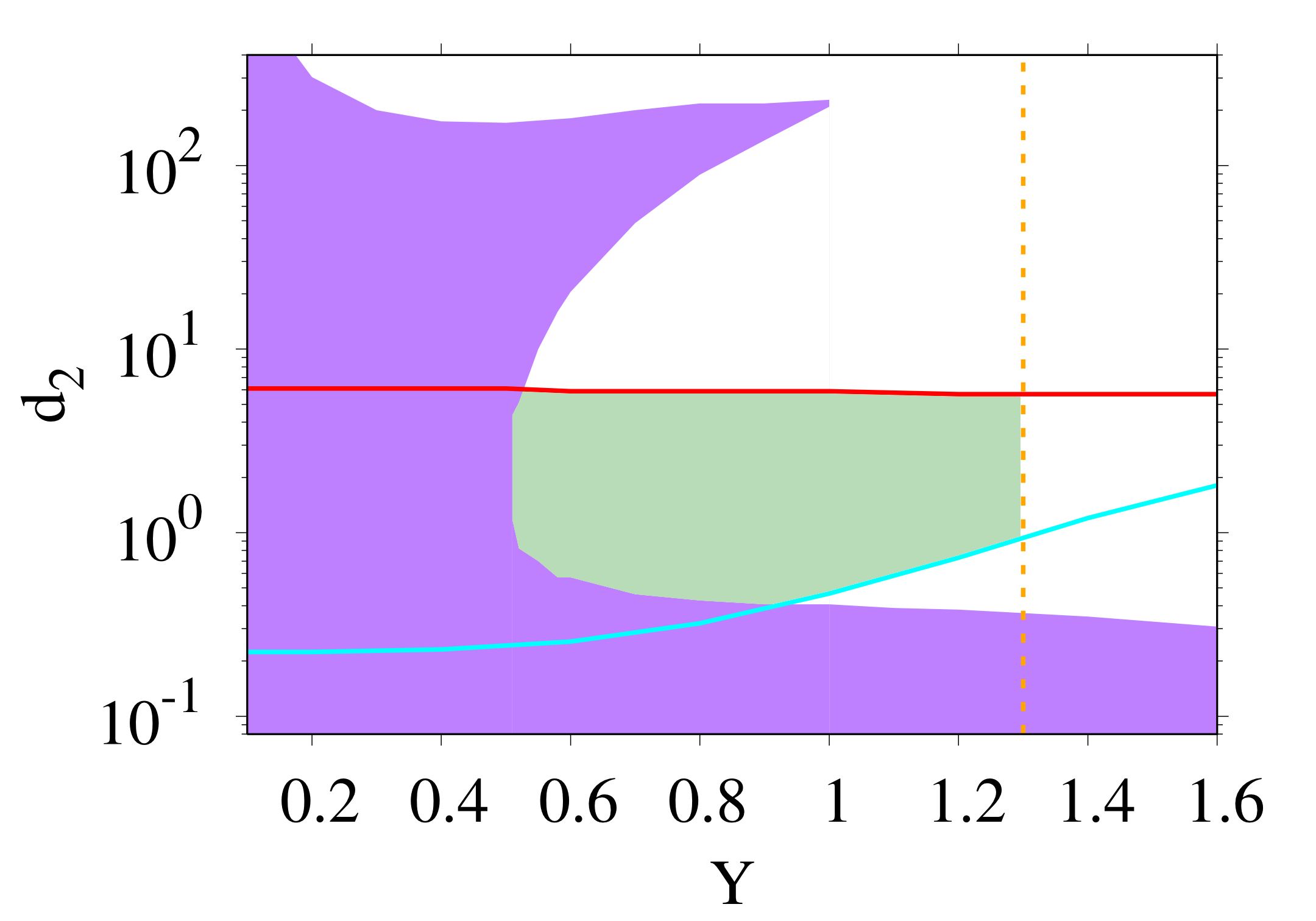}
 }

\caption{\small{
The allowed region for the CSF-2 model, with $\alpha = 0.05$ and $m^F_{\rm CDM} = 
m^S_{\rm CDM} + 200$ MeV. 
The region below the red horizontal line, above the cyan line, 
right to the overclosure patch, and left to the vertical line, is allowed.
}}
\label{fig:cdm-csf4}
\end{figure}

With these two possible candidates, there can be 
interconversions, $\bar\psi\psi \leftrightarrow S_2 S_2$, and the standard annihilations to 
bosonic and fermionic final states.  
The composition of the dark matter is shown in Fig.\ \ref{fig:cdm-csf1}, with $Y=0.7$ and scalar VEV $s_1=100$ 
GeV. In the left plot, we take 
$m^F_{\rm CDM} > m^S_{\rm CDM}$, and in the right plot, $m^F_{\rm CDM} < m^S_{\rm CDM}$,
the mass of the lighter one being fixed at 100 GeV. Note that if $m^S_{\rm CDM} > m^F_{\rm CDM}$, 
it quickly saturates the $\Omega h^2 = 0.12$ limit. If $m^F_{\rm CDM}$ is heavier of the two,
its contribution slowly rises with the mass difference $\Delta M$, 
and hits the relic density limit at $\Delta M \approx 450$ GeV. 
One may note that for the left panel of Fig.\ \ref{fig:cdm-csf1}, both $S_2$ and $\psi$ can be 
CDM components, while for the right panel, $S_2$ ceases to be a CDM component if the channel $S_2\to
\psi\bar\psi$ opens up. However, the overclosure bound is reached much before that.

The allowed parameter space for the CSF-2 model is shown in Fig.\ \ref{fig:cdm-csf2}, with $m^F_{\rm CDM} = 
m^S_{\rm CDM} + 200$ MeV, $\alpha = 0.05$ and $Y=0.7$. One finds that the plots display 
traits similar to both CS and RSF 
models, as expected. For example, the line showing the invisible decay constraint starts out vertically for 
small $d_2$ or $\delta_2$, like the RSF model, and then shows a rightward shift as found for the CS model. 
One may note the important role the triviality and unitarity lines play; they cut out a significant portion of the 
otherwise allowed parameter space. 

Fig.\ \ref{fig:cdm-csf3} shows the allowed region for the CDM-nucleon scattering cross-section, analogous to
Fig.\ \ref{fig:cdm-cs3}. Fig.\ \ref{fig:cdm-csf4}, similarly, shows the allowed region for the parameters of the potential,
analogous to Fig.\ \ref{fig:cdm-rsf4}. 

\section{Conclusion} \label{sec:conclude}

In this paper, we have studied the parameter space for several extensions of the SM that provide one or 
more cold dark matter candidates as well as a scalar at 96 GeV. The existence of the latter was only hinted by the 
CMS Collaboration, but if tagged with a CDM model, this provides further constraints on the parameter space, by 
reducing the number of free parameters in the Lagrangian. 

We analysed the parameter space of three models: (i) SM plus a complex scalar singlet (CS), with one of the 
scalars being the CDM, and the other giving rise to the 96 GeV resonance after mixing with the SM doublet; (ii) 
SM plus a real scalar singlet and a singlet vectorial fermion (RSF), with the fermion being the dark matter; and (iii) 
CS plus a singlet vectorial fermion (CSF), with both fermion and scalar being dark matter candidates. The theoretical 
constraints that were taken into account includes the stability of the potential (both triviality and boundedness),
the scattering unitarity (although it hardly differs from the triviality constraints), and the oblique parameters, 
the effect of the latter being negligible. The experimental constraints include those coming from the direct detection
of dark matter, the overclosure bound on the relic density ($\Omega h^2 < 0.12$), and the invisible decay width 
of the Higgs boson. We have assumed a thermalised nonrelativistic dark matter and used micrOMEGAs v5.0.8
to generate the relic density.  

At the same time, we have shown how the diphoton signal can be explained, either in some 
ultraviolet-complete theory, or in some effective theory framework. Combined with the CDM models, they 
provide a number of significant constraints on the parameter space of such models. 
In this paper, we collect and display for the first time, all such possible theoretical and experimental constraints 
on the allowed parameter space of these three models, extending the CDM mass to 1 TeV. 
Several interesting features emerge from the analysis, and they have been displayed in the previous section. 
One may enlist them once again here:

\begin{enumerate}

\item The triviality/unitarity constraints play a vital role in restricting the allowed parameter space for the 
CS and the CSF-2 models. In fact, the parameter space gets further squeezed if we assume the possible onset of a 
new physics at higher than 100 TeV. The severity of their effect depends on the singlet-doublet mixing angle 
$\alpha$.
If $\alpha$ is small enough, one finds that these constraints are always more powerful than those coming from direct 
detection. With larger values of $\alpha$, they become comparable. On the other hand, the constraint arising
out of the requirement of stability of the potential is significant only in the RSF model, or for larger values of
$\alpha$. 

\item The CS  and the CSF-2 models allow two narrow resonance regions, at approximately $m_h/2$ and 
$m_\chi/2$, and they become wider as the mixing angle increases. Thus, one may still have a sub-100 GeV scalar 
dark matter, which is not overly fine-tuned. However, $m_{\rm CDM} < m_\chi/2$ is ruled out from the 
CMS signal strength. 

\item The RSF model is much more tightly constrained than its CS or CSF counterparts. 
A large part of the parameter space allowed from the 
relic density bound is ruled out by triviality and unitarity limits. However, there exists a narrow window, 
which is one of the novel findings of this paper, and whose 
position depends on the singlet scalar VEV, while the width depends on the Yukawa coupling $Y$.
The ruled-out neck region, just above the window, appears because of the destructive interference between 
the $s$- and $t$-channel annihilation amplitudes, although a significant part of this window is truncated from the 
stability of potential.  
\end{enumerate}

While such a study might be interesting to the model builders as well as those looking for collider signatures 
of beyond-SM physics, one must be cautious in applying these bounds. The parameter space, even after 
allowing for the 96 GeV scalar, is complicated enough, and we have refrained from doing a complete scan over 
all the parameters. The allowed regions will shift for different values of the scalar VEV $s_1$, the CDM mass, 
or the splitting between the two CDM candidates for the CSF-2 model. However, we expect the general trends 
to remain qualitatively similar.


{\bf Acknowledgement:} AK acknowledges the Science and Engineering Research Board, Government of India, 
for support through the Grant EMR/2016/001306 and the DIA fellowship. 


\end{document}